\documentclass[journal]{IEEEtran}
\IEEEoverridecommandlockouts
\usepackage{amssymb, amsmath, amsfonts,  bm}
\usepackage[linesnumbered, ruled]{algorithm2e}
\usepackage{changepage}

\DeclareMathOperator*{\argmin}{\arg\!\min}

\usepackage[final]{graphicx}
\usepackage{graphics}
\usepackage{epstopdf}
\usepackage{float}
\usepackage[compress,nospace]{cite}
\usepackage{subfigure}
\usepackage{threeparttable}
\usepackage{fancyhdr} 
\usepackage[usenames, dvipsnames]{xcolor}
\usepackage{pgfplots, tikz}
\usepackage{tkz-berge}
\usetikzlibrary{plotmarks, patterns, arrows, automata, petri, topaths}
\usetikzlibrary{shapes.geometric}
\RequirePackage[l2tabu, orthodox]{nag}

\begin{document}

\title{Robust Calibration of Radio Interferometers in Non-Gaussian Environment}

\author{\IEEEauthorblockN{Virginie Ollier,
Mohammed Nabil El Korso,
R\'{e}my Boyer, \textit{Senior Member, IEEE},
Pascal Larzabal, \textit{Member, IEEE} and Marius Pesavento, \textit{Member, IEEE}}

\thanks{Virginie Ollier and Pascal Larzabal are with SATIE, UMR 8029, \'{E}cole Normale Sup\'{e}rieure de Cachan, Universit\'{e} Paris-Saclay, Cachan, France (e-mail: virginie.ollier@satie.ens-cachan.fr, pascal.larzabal@satie.ens-cachan.fr). Virginie Ollier and R\'{e}my Boyer are with L2S, UMR 8506, Universit\'{e} Paris-Sud, Universit\'{e} Paris-Saclay, Gif-sur-Yvette, France (e-mail: remy.boyer@l2s.centralesupelec.fr). Mohammed Nabil El Korso is with LEME, EA 4416, Universit\'{e} Paris-Ouest, Ville d'Avray, France (e-mail: m.elkorso@parisnanterre.fr). Marius Pesavento is with Communication Systems Group, Technische Universit\"{a}t Darmstadt, Darmstadt, Germany (e-mail: mpesa@nt.tu-darmstadt.de).

This work was supported by MAGELLAN (ANR-14-CE23-0004-01), ON FIRE project (Jeunes Chercheurs GDR-ISIS) and ANR ASTRID project MARGARITA.}
}

\maketitle
\begin{abstract}
The development of new phased array systems in radio astronomy, as the low frequency array (LOFAR) and the square kilometre array (SKA), formed of a large number of small and flexible elementary antennas, has led to significant challenges. Among them, model calibration is a crucial step in order to provide accurate and thus meaningful images and requires the estimation of all the perturbation effects introduced along the signal propagation path, for a specific source direction and antenna position. Usually, it is common to perform model calibration using the a priori knowledge regarding a small number of known strong calibrator sources but under the assumption of Gaussianity of the noise. Nevertheless, observations in the context of radio astronomy are known to be affected by the presence of outliers which are due to several causes, e.g., weak non-calibrator sources or man made radio frequency interferences. Consequently, the classical Gaussian noise assumption is violated leading to severe degradation in performances. In order to take into account the outlier effects, we assume that the noise follows a spherically invariant random distribution. Based on this modeling, a robust calibration algorithm is presented in this paper.
More precisely, this new scheme is based on the design of an iterative relaxed concentrated maximum likelihood estimation procedure which allows to obtain closed-form expressions for the unknown parameters with a reasonable computational cost. This is of importance as the number of estimated parameters depends on the number of antenna elements, which is large for the new generation of radio interferometers. Numerical simulations reveal that the proposed algorithm outperforms the state-of-the-art calibration techniques.

 \end{abstract}
 
 
\section{Introduction}

Radio astronomy aims to study radio emissions from the sky, in order to detect, identify new objects and observe known structures at higher resolution, in a specific electromagnetic spectrum \cite{thompson2008interferometry}. This fundamental thematic shines a new light on our universe, revealing more about its nature and history. 
In order  to carry out particularly sensitive observations in a large range of the spectrum, and to handle significant cosmological issues, largely distributed sensor arrays 
are currently being built or planned, such as the low frequency array (LOFAR) \cite{van2013lofar} and the square kilometre array (SKA) \cite{dewdney2009square}. They will notably be composed of a large number of relatively low-cost small antennas with wide field of view, resulting in a large collecting area and high resolution imaging. Nevertheless, to meet the theoretical optimal performances of such next generation radio interferometers, a plethora of signal processing challenges must be overcome, among them, calibration, data reduction and image synthesis \cite{wijnholds2008fundamental,rau2009advances,deVos2009lofar, van2013signal}.
These aspects are intertwined and must be dealt with to take advantage of the new advanced radio interferometers. As an example, lack of calibration has dramatic effects in the image reconstruction by causing severe distortions. In this paper, we focus on calibration, which involves the estimation of all unknown perturbation effects and represents a cornerstone of the imaging step \cite{mitchell2008real,wijnholds2014signal,wijnholds2010calibration}.

Array calibration aspects have been tackled for a few decades in the array processing community leading to a variety of calibration algorithms \cite{fuhrmann1994estimation,ng1996sensor,ng2009practical}. Such algorithms can be classified into two different approaches depending on the presence \cite{lo1987eigenstructure,ng1992array,ng1995active}, or the absence \cite{rockah1987array,weiss1989array,friedlander1991direction,wylie1994joint,qiong2003overview,flanagan2001array}, of one or more cooperative sources, named calibrator sources. In the radio astronomy context, calibration is commonly treated using the first approach as we have access to prior knowledge thanks to tables describing accuratly the position and flux of the brightest sources \cite{baars1977absolute}.

Following this methodology, the majority of proposed calibration schemes in radio interferometry are least squares-based approaches. The state-of-the-art consists in the so-called alternating least squares approach \cite{boonstra2003gain, wijnholds2009multisource,wijnholds2010fish,salvini2014fast}, which leads to statistically efficient algorithm under a Gaussian model, since the least squares estimator is equivalent to the maximum likelihood (ML) estimator in this case. On the other hand, expectation maximization (EM) \cite{dempster1977maximum, moon1996expectation,mclachlan2007algorithm} and EM-based algorithms, such as the space alternating generalized expectation maximization algorithm \cite{fessler1994space}, have been proposed in order to enhance the convergence rate of the least squares-based calibration algorithms \cite{yatawatta2009radio}. Nevertheless, the major drawback of these schemes is the Gaussianity assumption which is not realistic in the radio astronomy context. Specifically, the presence of outliers has multiple causes, among which i) the radio frequency interferers, which corrupt the observations and are not always perfectly filtered in practice \cite{raza2002spatial,van2005performance}, ii) the presence of unknown weak sources in the background \cite{yatawatta2014robust}, iii) the presence of some punctual events such as interference due to the Sun or due to strong sources in the sidelobes which can also randomly create outliers \cite{boonstra2005radio}. To the best of our knowledge, the proposed scheme in \cite{yatawatta2014robust}, represents the only alternative to the existing calibration algorithms based on a Gaussian noise model.

In \cite{yatawatta2014robust}, theoretical and experimental analyses have been conducted in order to demonstrate that the effect of outliers in the radio astronomy context can indeed be modeled by a non-Gaussian heavy-tailed distributed noise process. Nevertheless, the algorithm presented in \cite{yatawatta2014robust} has its own limits, since the noise is specifically modeled as a Student's t with independent identically distributed entries. To improve the robustness of the calibration, we propose, in this paper, a new scheme based on a broader class of distributions gathered under the so-called spherically invariant random noise modeling \cite{jay2002detection,yao2003spherically}, which includes the Student's t distribution. A spherically invariant random vector (SIRV) is described as the product of a positive random variable, named texture, and the so-called speckle component which is  Gaussian, resulting in a two-scale compound Gaussian distribution \cite{wang2006maximum}.
The flexibility of the SIRV modeling allows to consider non-Gaussian heavy-tailed distributed noise in the presence of outliers, but also to adaptively consider Gaussian noise in the extreme case when there are no outliers. Under the SIRV model, we estimate the unknown parameters iteratively based on a relaxed ML estimator, leading to closed-form expressions for the noise parameters while a block coordinate descent (BCD) algorithm \cite{friedman2007pathwise,hong2015unified} is designed to obtain the estimates of parameters of interest efficiently and at a low cost.

Finally, it is worth mentioning that the parametric model used in this paper to describe the perturbation effects is based on the so-called Jones matrices \cite{hamaker1996understanding,smirnov2011revisiting}. Such formalism describes in a flexible way the conversion of the incident electric field into voltages. Indeed, along its propagation path, the signal is affected by various effects and transformations which correspond to matrix multiplications in the mathematical Jones framework. Multiple distortion effects caused by the environment and/or the instruments can be easily incorporated into the model using an adequate parametrization of the Jones matrices. Such effects can represent, for example, the ionospheric phase delay resulting in angular shifts, the atmospheric distortions, the typical phase delay due to geometric pathlength difference, the voltage primary beam, the cross-leakage or also the electronic gains \cite{noordam1996measurement,noordam2010meqtrees}. For the above reasons and due to its flexibility \cite{thompson2008interferometry,yatawatta2009radio,noordam1996measurement,hamaker1996understanding,smirnov2011revisiting}, we adopt this parametric model. We make a distinction between the non-structured and the structured cases: in the first one, one total Jones matrix stands for all the effects along the full signal path while in the second case, we regard each physical effect separately thanks to individual Jones terms in a cumulative product. Thus, different corruptions are described by different kinds of Jones matrices.
 We emphasize that the proposed algorithm, entitled relaxed concentrated ML estimator, is a generic algorithm as it is based on a non-structured Jones matrices formulation as a first step. However, it can be adapted to various regimes 
 describing distinct calibration scenarios in which an array can operate \cite{lonsdale2004calibration}. In this paper, we consider the specific example of the direction dependent distortion regime with a compact set of antennas, which we refer to as the 3DC regime.
The array is therefore considered as a closely packed group of antennas but the array elements have a wide field of view. This is particularly well-adapted for calibration of compact arrays, typically a LOFAR station.
 
 The rest of the paper is organized as follows: in Section \ref{data}, we present the data model  in the context of radio astronomy, first with non-structured Jones matrices and thereafter, we study an example of structured Jones matrices for the 3DC calibration regime. In Section \ref{robust_estim}, we give an overview of the proposed robust ML estimator, based on spherically invariant random process (SIRP) noise modeling. An efficient estimation procedure of the distortions introduced on each signal propagation path is derived in Section \ref{non_structure}. Then, the algorithm is adapted to the case of structured Jones matrices in Section \ref{structure} for the 3DC calibration regime. Finally, we provide numerical simulations in Section \ref{simus} to assess the robustness of the approach and draw  our conclusions in Section \ref{ccl}.

In this paper, we use the following notation: symbols $\left( \cdot \right) ^{T}$, $\left( \cdot \right)^{\ast }$, $\left( \cdot \right) ^{H}$ denote, respectively, the transpose, the complex conjugate and the Hermitian transpose. The Kronecker product is represented by $\otimes$, $\mathrm{E}\{\cdot\}$ denotes the expectation operator, $\mathrm{bdiag}\{\cdot\}$ is the block-diagonal operator, whereas $\mathrm{diag}\{\cdot\}$ converts a vector into a diagonal matrix. The trace and determinant operators are, respectively, referred by $\mathrm{tr}\left\{ \cdot \right\} $ and $ |\cdot|$. The symbol $\mathbf{I}_{B}$ represents the $B \times B$ identity matrix, $\mathrm{vec}(\cdot)$ stacks the columns of a matrix on top of one another, $||\cdot||_F$ is the Frobenius norm, while $||\cdot||_2$ denotes the $l_2$ norm. Finally, $\Re\left\{ \cdot \right\} $ represents the real part 
and we note $j$ the complex number whose square equals $-1$.

\section{Data model}
\label{data}

\subsection{Case of non-structured Jones matrices}
 
Let us consider $M$ antennas with known locations that receive $D$ signals emitted by calibrator sources. Each antenna is dual polarized and composed of two receptors, in order to provide sensitivity to the two orthogonal
polarization directions $(x,y)$ of the incident electromagnetic plane wave. Consequently, the relation between the \textit{i}-th source emission and the
measured voltage at the \textit{p}-th antenna is given by \cite{hamaker1996understanding,KerThesis2012,smirnov2011revisiting}
\begin{equation}
\mathbf{v}_{i,p}(\boldsymbol{\theta}) = \mathbf{J}_{i,p}(\boldsymbol{\theta}) \mathbf{s}_{i}
\end{equation}
where $\mathbf{s}_{i} = [
  s_{i_{x}},s_{i_{y}}]^{T}$ is the incoming signal, $ \mathbf{v}_{i,p}(\boldsymbol{\theta}) = [
  v_{i,p_{x}}(\boldsymbol{\theta}),
v_{i,p_{y}}(\boldsymbol{\theta})]^{T}$ is the generated voltage with one output for each polarization direction and $\mathbf{J}_{i,p}(\boldsymbol{\theta})$ denotes the so-called $2 \times 2$ Jones matrix, parametrized by the unknown vector of interest $\boldsymbol{\theta}$. The Jones matrix models the array response and all the perturbations introduced along the path from the \textit{i}-th source to the \textit{p}-th sensor. Since each propagation path is particular,
we can associate a different Jones matrix with each source-antenna pair $(i,p)$, leading to a total number of $D M$ Jones matrices.
In this section, we  consider the non-structured case where no specific perturbation model is used to describe the physical mechanism behind each perturbation effect and the unknown elements correspond to the entries of all Jones matrices \cite{yatawatta2009radio,nunhokeeghost2015} (a structured example is given for 3DC calibration regime, in Section \ref{specific_regime}).

For each antenna pair, we compute the correlation of the output signals, resulting in the typical observations recorded by a radio interferometer.
The correlation between voltages is given, in the case of noise free measurements, for the $(p,q)$ antenna pair, by
\begin{align}
\label{ME}
\nonumber
\mathbf{V}_{pq}(\boldsymbol{\theta}) & =\mathrm{E}\left\{\left(\sum_{i=1}^{D}\mathbf{v}_{i,p}(\boldsymbol{\theta})\right)\left(\sum_{i=1}^{D}\mathbf{v}_{i,q}^{H}(\boldsymbol{\theta})\right)\right\} \\ &=
\sum_{i=1}^{D}\mathbf{J}_{i,p}(\boldsymbol{\theta})\mathbf{C}_{i}
\mathbf{J}_{i,q}^{H}(\boldsymbol{\theta}) \ \
\text{for} \ \ p<q, \ \ p,q \in \{1, \ldots, M\},
\end{align}
where the signals emitted by the sources are assumed uncorrelated and the $2 \times 2$ matrix $\mathbf{C}_{i}=\mathrm{E}\{\mathbf{s}_{i}\mathbf{s}_{i}^{H}\}$
 is known from prior knowledge.
Let us remark that autocorrelations are not considered as shown by the condition $p<q$ in (\ref{ME})
(this is a typical situation in the radio astronomy context where radio interferometric systems automatically flag the autocorrelations \cite{van2013lofar}).

Using the property $\mathrm{vec}(\mathbf{A B C}) = (\mathbf{C}^{T} \otimes \mathbf{A})\mathrm{vec}(\mathbf{B})$, we rewrite (\ref{ME}) as a $4 \times 1$ vector
\begin{equation}
\tilde{\mathbf{v}}_{pq}(\boldsymbol{\theta})=\mathrm{vec}\Big(\mathbf{V}_{pq}(\boldsymbol{\theta})\Big)
= \sum_{i=1}^{D}\mathbf{u}_{i,pq}(\boldsymbol{\theta})
\end{equation}
 in which $\mathbf{u}_{i,pq}(\boldsymbol{\theta})=\left(\mathbf{J}^{\ast}_{i,q}(\boldsymbol{\theta})\otimes
\mathbf{J}_{i,p}(\boldsymbol{\theta})\right)\mathbf{c}_{i}$ with $\mathbf{c}_{i}=\mathrm{vec}(\mathbf{C}_{i})$.
We stack all the noisy measurements within a full vector $\mathbf{x}=\left[\mathbf{v}^{T}_{12},
\mathbf{v}^{T}_{13},
\ldots,
\mathbf{v}^{T}_{(M-1)M}\right]^{T}$ $\in\mathbb{C}^{4 B \times 1}$, where $B=\frac{M(M-1)}{2}$ denotes the total number of antenna pairs and $\mathbf{v}_{pq}=\tilde{\mathbf{v}}_{pq}(\boldsymbol{\theta})+\mathbf{n}_{pq}$ with $\mathbf{n}_{pq}$ the noise sample at a specific antenna pair. Specifically, $\mathbf{x}$ reads
\begin{equation}
\label{express_x}
\mathbf{x}
=\sum_{i=1}^{D}\mathbf{u}_{i}(\boldsymbol{\theta})+\mathbf{n}
\end{equation}
in which  $\mathbf{u}_{i}(\boldsymbol{\theta})=
  \left[\mathbf{u}^{T}_{i,12}(\boldsymbol{\theta}),
\mathbf{u}^{T}_{i,13}(\boldsymbol{\theta}),
\ldots,
\mathbf{u}^{T}_{i,(M-1)M}(\boldsymbol{\theta})\right]^{T}$ and $\mathbf{n}=\left[\mathbf{n}^{T}_{12},
\mathbf{n}^{T}_{13},
\ldots,
\mathbf{n}^{T}_{(M-1)M}\right]^{T}$ is the full noise vector which accounts for Gaussian noise, but also the presence of outliers in our data. Therefore, the noise can no longer be considered Gaussian and a robust calibration method is required. 
 To investigate non-Gaussian noise modeling and encompass a broad range of noise distributions, we propose to adopt the SIRP noise model \cite{jay2002detection,yao2003spherically}. Specifically, the noise at each antenna pair is assumed to be generated as
\begin{equation}
\label{sirp}
\mathbf{n}_{pq} = \sqrt{\tau_{pq}} \  \mathbf{g}_{pq},
\end{equation}
where the positive real random variable $\tau_{pq}$ is referred to as texture, whereas the complex speckle component $\mathbf{g}_{pq}$ follows a zero-mean Gaussian distribution\footnotemark, i.e.,
\begin{equation}
\label{constraint}
\mathbf{g}_{pq} \sim \mathcal{CN}(\mathbf{0},\boldsymbol{\Omega}).
\end{equation}
In order to remove scaling ambiguities, we impose $\mathrm{tr}\left\{\boldsymbol{\Omega} \right\}=1$. Note that the choice of this constraint is arbitrary and does not affect the estimates of interest as argued in \cite{zhang2016mimo}.

\footnotetext{Let us note that it is possible to consider a different covariance matrix $\boldsymbol{\Omega}_{pq}$ for each speckle component $\mathbf{g}_{pq}$ in (\ref{constraint}). In this case, the proposed algorithm requires a few modifications and the corresponding expressions are presented in Appendix A.
}

In this section, we adopted the non-structured Jones matrices formulation which is relevant in the radio astronomical context \cite{yatawatta2009radio,nunhokeeghost2015}. In this case, there is no need to specify the full propagation path, avoiding misspecification in the model. Besides, it is highly flexible and can be adapted to different scenarios \cite{lonsdale2004calibration}. In the following, we present the direction dependent distortion regime with a compact set of antennas, named 3DC regime.

\subsection{Specific case of the 3DC calibration regime}
\label{specific_regime}

  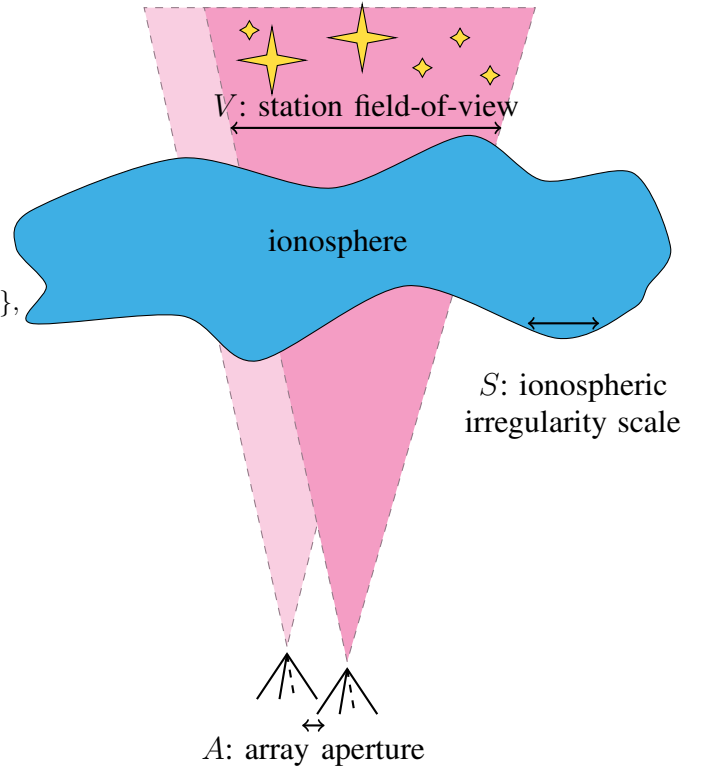
\begin{figure}
 \centering
 \begin{tikzpicture}
        \coordinate (f1) at (0.4,0.7);
        \coordinate (f2) at (-1.5,9.2);
        \coordinate (f3) at (2.5,9.2);
        \coordinate (f4) at (1.2,0.5);
        \coordinate (f5) at (-0.7,9.2);
        \coordinate (f6) at (3.7, 9.2);
        \coordinate (i1) at (-3, 5);
        \coordinate (i2) at (-2.8, 5.5);
        \coordinate (i3) at (-3.2, 6);
        \coordinate (i4) at (-3, 6.5); 
        \coordinate (i5) at (-1, 7.2);
        \coordinate (i51) at (1, 6.8);
        \coordinate (i6) at (2.8, 7.5);
        \coordinate (i7) at (3.8, 6.9);
        \coordinate (i8) at (5, 7); 
        \coordinate (i9) at (5.5, 6);
        \coordinate (i10) at (5.2, 5.5);
        \coordinate (i11) at (5, 5.2);
        \coordinate (i111) at (4, 4.8);
        \coordinate (i12) at (2, 5.5);
        \coordinate (i121) at (0, 4.5);
        \coordinate (i13) at (-1, 5.1);
        \coordinate (t4) at (2.5, 3);
        \coordinate (t3) at (4.2, 3.9);
        \coordinate (e1) at (1.4, 8.8);
        \coordinate (e2) at (0.2, 8.5);
            \coordinate (e3) at (2.2, 8.4);
        \coordinate (e4) at (-0.1, 8.9);
            \coordinate (e5) at (2.7, 8.8);
        \coordinate (e6) at (3.1, 8.3);
        \large
            \draw[color=black,thick] (0,0) -- +(0.4,0.6) -- +(0.8, 0); 
            \draw[color=black,thick] (0.4,0.6) -- +(-0.1,-0.6);
            \draw[color=black,thick,dashed] (0.4,0.6) -- +(+0.1,-0.6);
            \draw[color=black,thick] (0.8,-0.2) -- +(0.4,0.6) -- +(0.8, 0); 
            \draw[color=black,thick] (1.2,0.4) -- +(-0.1,-0.6);
            \draw[color=black,thick,dashed] (1.2,0.4) -- +(+0.1,-0.6);
            \fill[color=Lavender, opacity=0.5] (f1) -- (f2) -- (f3) -- cycle;
            \draw[color=Lavender!50!black,dashed] (f1) -- (f2) -- (f3) -- cycle;
            \fill[color=Lavender] (f4) -- (f5) -- (f6) -- cycle;
            \draw[color=Lavender!50!black,dashed] (f4) -- (f5) -- (f6) -- cycle;
            \fill[color=CornflowerBlue] plot [smooth cycle] coordinates {(i1) (i2) (i3) (i4) (i5) (i51) (i6) (i7) (i8) (i9) (i10) (i11) (i111) (i12) (i121) (i13)};
            \draw[color=black] plot [smooth cycle] coordinates {(i1) (i2) (i3) (i4) (i5) (i51) (i6) (i7) (i8) (i9) (i10) (i11) (i111) (i12) (i121) (i13)};
            \draw[color=black,thick, <->] (0.6, -0.35) -- (0.9,-0.35) node[midway, below] {$A$: array aperture};
            \draw[color=black,thick, <->] (-0.35,7.6) -- (3.25, 7.6) node[midway, above] {$V$: station field-of-view};
            \draw[color=black,thick, <->] (3.6,5) -- (4.55,5); 
            \draw (0,6.4) node[below right] {ionosphere} node{};
            \draw (t3) node[text width=3.3cm, text centered] {$S$: ionospheric irregularity scale} node{};
            \draw (e1) node[star, star points=4, star point ratio=4.5, draw,inner sep=2pt, fill=Goldenrod]{ };
            \draw (e2) node[star, star points=4, star point ratio=4.5, draw,inner sep=2pt, fill=Goldenrod]{ };
            \draw (e3) node[star, star points=4, star point ratio=1.95, draw,inner sep=1.3pt, fill=Goldenrod]{ };
            \draw (e4) node[star, star points=4, star point ratio=1.95, draw,inner sep=1.3pt, fill=Goldenrod]{ };
            \draw (e5) node[star, star points=4, star point ratio=1.95, draw,inner sep=1.3pt, fill=Goldenrod]{ };
            \draw (e6) node[star, star points=4, star point ratio=1.95, draw,inner sep=1.3pt, fill=Goldenrod]{ };
\end{tikzpicture}
 \caption{3DC calibration regime, for which $V\gg S$ and $S\gg A$. All receiving elements in the station see the same ionosphere part but, due to their wide field of view, a multitude of sources are visible and perturbations are highly direction dependent.}
 \label{regime3}
\end{figure}

For a specific propagation path, from the \textit{i}-th source to the \textit{p}-th antenna, the global Jones matrix $\mathbf{J}_{i,p}$ accounts for multiple effects which can be described explicitly. Indeed, each global matrix can be decomposed into individual Jones terms which stand for specific physical effects \cite{noordam1996measurement,smirnov2011revisiting}. This way, instead of estimating entries of all Jones matrices as done in the non-structured case, we will estimate physically meaningful parameters, thus reducing the total number of parameters to estimate. Introducing structured Jones matrices  can be done in the context of calibration scenarios\cite{lonsdale2004calibration,wijnholds2010calibration}. In what follows, we target one specific commonly used calibration regime that we call 3DC calibration regime, described in Fig. \ref{regime3}, which is well adapted for calibration of LOFAR on station level \cite{wijnholds2010fish}, and also for stations of the future SKA radio interferometer \cite{wijnholds2010calibration}. In this scenario, direction dependent distortions play a significant role since individual elements in the array have a wide field of view. Indeed, this implies different propagation conditions towards distinct sources in the field of view.
However, the array being relatively compact, made of similar elements, some effects might be the same for all antennas.
In the following, we introduce a particular sequence of Jones matrices with specific parametrizations, in the context of 3DC calibration regime \cite{yatawatta2012reduced}
\begin{equation}
 \label{model_regime3}
\mathbf{J}_{i,p}(\boldsymbol{\theta}^{\mathrm{3DC}}_{i,p})=\mathbf{G}_p(\mathbf{g}_p)\mathbf{H}_{i,p}\mathbf{Z}_{i,p}(\boldsymbol{\alpha}_i)\mathbf{F}_{i}(\vartheta_{i}) 
\end{equation}
for $i \in \{1, \hdots, D\}$, $p \in \{1, \hdots, M\}$ and $\boldsymbol{\theta}^{\mathrm{3DC}}_{i,p}=[\vartheta_{i},\mathbf{g}^T_p,\boldsymbol{\alpha}^T_i]^T$. We note $\mathbf{H}_{i,p}$ the only assumed known matrix thanks to electromagnetic simulations in terms of antenna response and a priori knowledge given by calibrator source and antenna positions \cite{yatawatta2012reduced,kazemi2015blind,noordam1996measurement,smirnov2011revisiting}, whereas the remaining matrices are explained in the following items.

$\bullet$ \textit{Ionospheric effects} :\\
Propagation through the ionosphere, the outer layer of the earth's atmosphere, introduces propagation delay on the signal which is affected by spatially variable fluctuations. 
If these perturbation effects are not corrected for, the sources may appear shifted from their intrinsic positions \cite{wijnholds2010calibration,van2007self}. In the case of a compact array, the ionospheric delay matrix is in fact a scalar direction-dependent phase given by
  \begin{equation}
  \label{form_Z}
  \mathbf{Z}_{i,p}(\boldsymbol{\alpha}_i)=\exp\Big\{j\varphi_{i,p}\Big\}\mathbf{I}_2
  \end{equation}
  in which $\varphi_{i,p}=\eta_i u_p+\zeta_i v_p$ where $\boldsymbol{\alpha}_i=[\eta_i,\zeta_i]^T$ is the vector of unknown offsets resulting in a shift of the \textit{i}-th source direction and $\mathbf{r}_p=[u_p,v_p]^T$ is the vector of known antenna position in units of wavelength. 
  
On top of that, passing through the ionosphere is associated with a rotation of the polarisation plane of each signal source around the line of sight. We call it   the ionospheric Faraday rotation matrix $\mathbf{F}_{i}(\vartheta_{i})$ and write it as
\begin{equation}
\mathbf{F}_{i}(\vartheta_{i}) = \begin{bmatrix}
    \cos(\vartheta_{i}) & -\sin(\vartheta_{i}) \\
   \sin(\vartheta_{i}) & \cos(\vartheta_{i})
\end{bmatrix}
\end{equation}
where $\vartheta_i$ is the unknown Faraday rotation angle, assumed identical for all antennas, since the array has a limited spatial extent \cite{noordam1996measurement}.
  
  $\bullet$ \textit{Instrumental effects} :\\
  Individual antennas are described by electronic complex gains which appear in
  $\mathbf{G}_p(\mathbf{g}_p)=\mathrm{diag}\{\mathbf{g}_p\}$ with $\mathbf{g}_p$ the unknown electronic gain vector.

  Therefore, in this specific structured case, the physical model parameters in (\ref{model_regime3}) are collected in the vector $\boldsymbol{\varepsilon}^{\mathrm{3DC}}=\mathbf{P}[\boldsymbol{\theta}^{\mathrm{3DC}^T}_{1,1},\boldsymbol{\theta}^{\mathrm{3DC}^T}_{1,2},\hdots,\boldsymbol{\theta}^{\mathrm{3DC}^T}_{D,M}]^T$ where $\mathbf{P}$ is an appropriate rearrangement matrix such that $\boldsymbol{\varepsilon}^{\mathrm{3DC}}=[\vartheta_{1}, \hdots,\vartheta_{D}, \mathbf{g}^T_1, \hdots, \mathbf{g}^T_M, \boldsymbol{\alpha}^T_1 , \hdots, \boldsymbol{\alpha}^T_D]^T$.

\section{Robust calibration estimator}
\label{robust_estim}

This section is devoted to the design of a robust calibration estimator based on the model (\ref{express_x}). As it can be seen from (\ref{sirp}), one has to specify the probability density function (pdf) of each texture parameter $\tau_{pq}$ in order to obtain the exact ML estimates. Nevertheless, in pratical scenarios, such prior knowledge is not available. Consequently, our idea is to make use of a relaxed version of the exact model, i.e., we assume deterministic but unknown texture realizations in the estimation process \cite{conte2002recursive,pascal2008covariance}. This ensures more flexibility in our algorithm as the texture distribution is not precisely described and avoids any possible model misspecification, which is consistent with our motivation to design a broad robust estimator w.r.t.~the presence of outliers. On the other hand, we adopt here an iterative procedure in order to reduce the computational cost. In doing so, the proposed algorithm sequentially updates each block of unknown parameters while fixing the remaining parameters. This leads to a relaxed concentrated ML based calibration estimator for which the expression of the likelihood function, when independency is assumed between measurements,  is written as
\begin{align}
\label{likelihood}
\nonumber
& f(\mathbf{x}|\boldsymbol{\theta },
\boldsymbol{\tau}, \boldsymbol{\Omega})= \\  &
\prod_{pq}\frac{1}{| \pi
\tau_{pq} \boldsymbol{\Omega} |} \exp
\left\{-\frac{1}{\tau_{pq}}\mathbf{a}_{pq}^{H}(\boldsymbol{\theta})
\boldsymbol{\Omega}^{-1}\mathbf{a}_{pq}(\boldsymbol{\theta})\right\},
\end{align}
where the vector composed of all texture realizations is $\boldsymbol{\tau}=[\tau_{12},\tau_{13},\ldots,\tau_{(M-1)M}]^{T}$ and $\mathbf{a}_{pq}(\boldsymbol{\theta}) = \mathbf{v}_{pq}- \tilde{\mathbf{v}}_{pq}(\boldsymbol{\theta})$.
Consequently, the log-likelihood function reads
\begin{align}
\label{loglikelihood}
\nonumber
 & \log f(\mathbf{x}|\boldsymbol{\theta },
\boldsymbol{\tau}, \boldsymbol{\Omega}) =
-4B\log\pi \\  &
-4\sum_{pq}\log\tau_{pq}-
B\log |\boldsymbol{\Omega}|-\sum_{pq}\frac{1}{\tau_{pq}}\mathbf{a}_{pq}^{H}(\boldsymbol{\theta})
\boldsymbol{\Omega}^{-1}\mathbf{a}_{pq}(\boldsymbol{\theta}).
\end{align}

In the following, we present the sequential updates of each block of unknown parameters, namely, $\boldsymbol{\tau}$,  $\boldsymbol{\Omega}$ and $\boldsymbol{\theta}$, following the methodology as in \cite{zhang2015maximum,ollier2016relaxed}.

\textbf{\textit{1) Derivation of $\hat{\tau}_{pq}$}}:
Taking the derivative of the log-likelihood function in (\ref{loglikelihood}) w.r.t.~$\tau_{pq}$
leads to the following texture estimate
\begin{equation}
\label{tauExp} \hat{\tau}_{pq} = \frac{1}{4} \mathbf{a}_{pq}^{H}(\boldsymbol{\theta})
\boldsymbol{\Omega}^{-1} \mathbf{a}_{pq}(\boldsymbol{\theta}).
\end{equation}

\textbf{\textit{2) Derivation of $\hat{\boldsymbol{\Omega}}$}}:
The derivative of the log-likelihood function w.r.t.~the element
$[\boldsymbol{\Omega}]_{k,l}$ of the speckle covariance matrix, using classical differential properties \cite[p.~2741]{hjorungnes2007complex}, leads to
\begin{equation} 
\label{OmegaDerive}
-B\mathrm{tr}\left\{\hat{\boldsymbol{\Omega}}^{-1}\mathbf{e}_{k}\mathbf{e}_{l}^{T}\right\}+
\sum_{pq}\frac{1}{\tau_{pq}}\mathbf{a}_{pq}^{H}(\boldsymbol{\theta})
\hat{\boldsymbol{\Omega}}^{-1}\mathbf{e}_{k}\mathbf{e}_{l}^{T}\hat{\boldsymbol{\Omega}}^{-1}\mathbf{a}_{pq}(\boldsymbol{\theta})=0
\end{equation}
where the vector $\mathbf{e}_{k}$ contains zeros except at the
$k$-th position which is equal to unity. The permutation property of the trace operator enables to rewrite (\ref{OmegaDerive}) as
\begin{equation}
-B\mathbf{e}_{l}^{T}\hat{\boldsymbol{\Omega}}^{-1}\mathbf{e}_{k}+
\sum_{pq}\frac{1}{\tau_{pq}}\mathbf{e}_{l}^{T}\hat{\boldsymbol{\Omega}}^{-1}\mathbf{a}_{pq}(\boldsymbol{\theta})\mathbf{a}_{pq}^{H}(\boldsymbol{\theta})
\hat{\boldsymbol{\Omega}}^{-1}\mathbf{e}_{k}=0.
\end{equation}
This finally leads to the following estimate of the speckle covariance matrix 
\begin{equation}
\label{OmegaEstimBefore}
\hat{\boldsymbol{\Omega}}=\frac{1}{B}\sum_{pq}\frac{1}{\tau_{pq}}\mathbf{a}_{pq}(\boldsymbol{\theta})\mathbf{a}_{pq}^{H}(\boldsymbol{\theta}).
\end{equation}
Inserting (\ref{tauExp}) into (\ref{OmegaEstimBefore}), we obtain
\begin{equation}
\label{OmegaEstim} \hat{\boldsymbol{\Omega}}^{h+1} = \frac{4}{B}\sum_{pq}
 \frac{\mathbf{a}_{pq}(\boldsymbol{\theta})
\mathbf{a}_{pq}^{H}(\boldsymbol{\theta})}{\mathbf{a}_{pq}^{H}(\boldsymbol{\theta})\Big(\hat{\boldsymbol{\Omega}}^{h}\Big)^{-1}\mathbf{a}_{pq}(\boldsymbol{\theta})}
\end{equation}
where $h$ denotes the \textit{h}-th iteration. Due to the introduced constraint, we normalize the estimate of $\boldsymbol{\Omega}$ by its trace, as follows
\begin{equation}
\label{trace}
\hat{\boldsymbol{\Omega}}^{h+1}=\frac{\hat{\boldsymbol{\Omega}}^{h+1}}{\mathrm{tr}\left\{\hat{\boldsymbol{\Omega}}^{h+1}\right\}}.
\end{equation}

\textbf{\textit{3) Estimation of $\hat{\boldsymbol{\theta}}$}}:
For given $\boldsymbol{\Omega}$ and $\boldsymbol{\tau}$, estimating $\hat{\boldsymbol{\theta}}$ is equivalent to the following minimization problem
\begin{equation}
\label{optimtheta}
\hat{\boldsymbol{\theta }}=\argmin_{\boldsymbol{\theta } }
\left\{\sum_{pq}\frac{1}{\tau_{pq}}\mathbf{a}_{pq}^{H}(\boldsymbol{\theta})
\boldsymbol{\Omega}^{-1}\mathbf{a}_{pq}(\boldsymbol{\theta})\right\}.
\end{equation}
 In the following, we aim to reduce the computational cost of the minimization procedure in (\ref{optimtheta}) by use of the EM algorithm. For generality, we first adopt the non-structured Jones matrix formulation, which can also be specified depending on the scenario, as shown in Section \ref{structure}.

\section{Estimation of $\hat{\boldsymbol{\theta}}$ for non-structured Jones matrices}
\label{non_structure}

Estimating directly the entries of the Jones matrices avoids specifying any particular physical model, leading to a calibration algorithm which is less sensitive to model errors in comparison with algorithms based on the structured case. Due to the possible large size of $\boldsymbol{\theta}$, a multi-dimensional parameter search needs to be carried out to solve the optimization problem in (\ref{optimtheta}) which requires significant computation time. To reduce this complexity, we make use of the EM algorithm.

The essence of this algorithm relies on a proper parameter vector partitioning as well as an adequate choice of the so-called complete data. 
As mentioned above, the parameters of interest $
\boldsymbol{\theta}$ represent the entries of all Jones matrices. Consequently, it is natural to consider the following partition
\begin{equation}
\label{express_theta}
\boldsymbol{\theta} = [\boldsymbol{\theta}_{1}^{T}, \ldots,
\boldsymbol{\theta}_{D}^{T}]^{T} =
[\boldsymbol{\theta}_{1,1}^{T}, \ldots,
\boldsymbol{\theta}_{1,M}^{T}, \ldots,
\boldsymbol{\theta}_{D,1}^{T},
\ldots,\boldsymbol{\theta}_{D,M}^{T} ]^{T},
\end{equation}
for which the vector $\boldsymbol{\theta}_{i,p}$ $\in\mathbb{R}^{8 \times 1}$ is the parametrization of the path from the \textit{i}-th calibrator source to the \textit{p}-th sensor, i.e.,
$\mathbf{J}_{i,p}(\boldsymbol{\theta})=\mathbf{J}_{i,p}(\boldsymbol{\theta}_{i,p})$.

\subsection{Use of the EM algorithm to solve (\ref{optimtheta})}
The EM algorithm \cite{dempster1977maximum,moon1996expectation, mclachlan2007algorithm} enables to compute the ML estimates and reduce the computational cost, via the iteration of two steps. The first one is the E-step which reduces, in our scenario, to the computation of the conditional expectation of the complete data given the observed data and the current estimate of parameters \cite{feder1988parameter,yatawatta2009radio}. Afterwards, the log-likelihood function of this conditional distribution is maximized in the M-step. Therefore, this last step consists in an optimization process which can be performed numerically, for instance with the Levenberg-Marquardt (LM) algorithm \cite{madsen1999methods,nocedal2006numerical,gavin2011levenberg}, or analytically if closed-form expressions are available. As we show in the following, the optimization step is carried out w.r.t.~to $\boldsymbol{\theta}_{i}$ $\in\mathbb{C}^{4 M \times 1}$ instead of  $\boldsymbol{\theta}$ $\in\mathbb{C}^{4 D M \times 1}$. Therefore, the global multiple source estimation problem is reduced to multiple single source sub-problems.

\textbf{\textit{1) E-step}}:
For the \textit{i}-th source, we introduce the so-called complete data vector $\mathbf{w}_{i}$ such that 
\begin{equation}
\mathbf{x}=\sum_{i=1}^{D}\mathbf{w}_{i}
\end{equation}
with $\mathbf{w}_{i} =
\mathbf{u}_{i}(\boldsymbol{\theta}_{i})+\mathbf{n}_{i}$ and $\mathbf{n}=\sum_{i=1}^{D}\mathbf{n}_{i}$, in which $\mathbf{n}_{i} \sim \mathcal{CN}(\mathbf{0},\beta_{i}\boldsymbol{\Psi})$. We have $\sum_{i=1}^{D}\beta_{i} =
1$ and  $\boldsymbol{\Psi}$ is the covariance matrix of $\mathbf{n}$.
Since $\mathbf{n}_{pq} \sim \mathcal{CN}(\mathbf{0},\tau_{pq} \boldsymbol{\Omega})$ and with the independence property, we obtain the following block-diagonal expression for $\boldsymbol{\Psi}$ 
\begin{equation}
\label{formPsi}  \boldsymbol{\Psi}  =\mathrm{bdiag}\{
    \tau_{12} \boldsymbol{\Omega} ,
    \ldots, \tau_{(M-1)M}\boldsymbol{\Omega}\}.
\end{equation}
Let us note $\mathbf{w} =[\mathbf{w}_{1}^{T}, \ldots, \mathbf{w}_{D}^{T}]^{T}$ the complete data vector, whose covariance matrix, denoted as $\boldsymbol{\Xi}$, has the following form
\begin{equation} 
\boldsymbol{\Xi}=\mathrm{bdiag}\{
    \beta_{1} \boldsymbol{\Psi}, \ldots, \beta_{D}\boldsymbol{\Psi}\}.
    \end{equation}

With \cite[p.~36]{anderson1958introduction}, and after some calculus, the expression of the conditional expectation is given by 
\begin{align}
\label{wRealEstim}  \nonumber \hat{\mathbf{w}}_{i}
& = \mathrm{E}\{\mathbf{w}_i|\mathbf{x};
\boldsymbol{\theta}, \boldsymbol{\tau}, \boldsymbol{\Omega}\} \\ & =\mathbf{u}_{i}(\boldsymbol{\theta}_{i}) +
\beta_{i}\left(\mathbf{x}-\sum_{l=1}^{D}\mathbf{u}_{l}(\boldsymbol{\theta}_{l})\right).
\end{align}

\textbf{\textit{2) M-step}}:
The goal of this step is to estimate $\boldsymbol{\theta}_{i}$. Once $\hat{\mathbf{w}}_{i}$ are computed for $ i \in
\{1, \ldots, D\}$, the estimated complete data vector $\hat{\mathbf{w}}$ can be evaluated. The M-step is an optimization problem based on the following likelihood function where $\mathbf{w}_{i}$ are independent 
\begin{align}
& \nonumber  f(\hat{\mathbf{w}}|\boldsymbol{\theta },\boldsymbol{\tau},\boldsymbol{\Omega})  = \\ &
\nonumber
\frac{1}{|\pi \boldsymbol{\Xi}|} \exp
\left\{-\Big(\hat{\mathbf{w}}-\mathbf{u}(\boldsymbol{\theta})\Big)^{H}\boldsymbol{\Xi}^{-1}\Big(\hat{\mathbf{w}}-\mathbf{u}(\boldsymbol{\theta})\Big)\right\}= \\ &
\prod_{i=1}^{D}\frac{1}{|\pi
\beta_{i}\boldsymbol{\Psi}|} 
\exp
\left\{-\Big(\hat{\mathbf{w}}_{i}-\mathbf{u}_{i}(\boldsymbol{\theta}_{i})\Big)^{H}(\beta_{i}\boldsymbol{\Psi})^{-1}\Big(\hat{\mathbf{w}}_{i}-\mathbf{u}_{i}(\boldsymbol{\theta}_{i})\Big)\right\}.
\end{align}
To obtain an estimation of $\boldsymbol{\theta}_{i}$, we need to minimize the following cost function
\begin{equation}
\label{Mstep_i}
\phi_{i}(\boldsymbol{\theta}_{i}) = \Big(\hat{\mathbf{w}}_{i}-\mathbf{u}_{i}(\boldsymbol{\theta}_{i})\Big)^{H}(\beta_{i}\boldsymbol{\Psi})^{-1}\Big(\hat{\mathbf{w}}_{i}-\mathbf{u}_{i}(\boldsymbol{\theta}_{i})\Big).
\end{equation}

To decrease even more the complexity cost of the proposed robust calibration scheme, we use the BCD algorithm \cite{friedman2007pathwise,hong2015unified} in the M-step. Consequently, we obtain analytical solutions for each single source sub-problems in (\ref{Mstep_i}), as shown below.

\subsection{Use of the BCD algorithm to minimize (\ref{Mstep_i})}

In (\ref{Mstep_i}), the optimization is performed w.r.t. $\boldsymbol{\theta}_{i}$. However, this unknown parameter vector can be partitioned according to the antennas, as expressed in (\ref{express_theta}). In the following, we perform the optimization of the cost function w.r.t. each $\boldsymbol{\theta}_{i,p}$ (\textit{p}-th antenna), given the current estimates of all the other $\boldsymbol{\theta}_{i,q}$ with $q\neq p$. This leads to a closed-form expression of $\hat{\boldsymbol{\theta}}_{i,p}$ as function of $\boldsymbol{\theta}_{i,q}$ for $q\neq p$ and the optimization process is repeated for each component vector $\boldsymbol{\theta}_{i,p}$ for $p \in \{1,\hdots,M\}$.

If we want to minimize (\ref{Mstep_i}) w.r.t. block-coordinate vector $\boldsymbol{\theta}_{i,p}$, we notice that only a subset of $\mathbf{u}_i(\boldsymbol{\theta}_i )$ is actually dependent on $\boldsymbol{\theta}_{i,p}$, i.e., $\{\mathbf{u}_{i,pq}\} \ \text{for} \ \ q > p, \ \ q \in \{1, \ldots, M\}$ and $\{\mathbf{u}_{i,qp}\} \ \text{for} \ \ q<p, \ \ q \in \{1, \ldots, M\}$. Therefore, (\ref{Mstep_i}) reads
\begin{align}
\label{Mstep_i_p}
\nonumber
& \phi_{i}(\boldsymbol{\theta}_{i,p})   = \\ & \nonumber \sum_{\substack {q=1\\ q>p}}^{M} \Big(\mathbf{w}_{i,pq} - \mathbf{u}_{i,pq}(\boldsymbol{\theta}_{i,p})\Big)^H
(\beta_i \tau_{pq} \boldsymbol{\Omega})^{-1} 
\Big(\mathbf{w}_{i,pq} - \mathbf{u}_{i,pq}(\boldsymbol{\theta}_{i,p})\Big)+ \\ &
\nonumber
\sum_{\substack {q=1 \\ q<p}}^{M} \Big(\mathbf{w}_{i,qp} - \mathbf{u}_{i,qp}(\boldsymbol{\theta}_{i,p})\Big)^H
(\beta_i \tau_{qp} \boldsymbol{\Omega})^{-1} 
\Big(\mathbf{w}_{i,qp} - \mathbf{u}_{i,qp}(\boldsymbol{\theta}_{i,p})\Big)+ \\ &
\mathrm{Constant}.
\end{align}
By $\mathrm{Constant}$, we mean the expressions independent of $\boldsymbol{\theta}_{i,p}$.
We show in Appendix B that it is possible to write $\mathbf{u}_{i,pq}$ directly as a function of $\boldsymbol{\theta}_{i,p}$, i.e., 
\begin{equation}
\label{form_pq}
\mathbf{u}_{i,pq}(\boldsymbol{\theta}_{i,p})=\boldsymbol{\Sigma}_{i,q}\boldsymbol{\theta}_{i,p}.
\end{equation}
In the same way, we have 
\begin{equation}
\label{form_qp}
\mathbf{u}_{i,qp}(\boldsymbol{\theta}_{i,p})=\boldsymbol{\Upsilon}_{i,q}\boldsymbol{\theta}^{\ast}_{i,p}.
\end{equation}
Notation and calculations being introduced in Appendix B, we only present here the results obtained, i.e., the expression of the estimated entries of the Jones matrix associated with the path from the \textit{i}-th calibrator source to the \textit{p}-th sensor which is given by
\begin{equation}
\label{theta_i_p_general}
\hat{\boldsymbol{\theta}}_{i,p} =\begin{small}
\left\lbrace
\begin{array}{lll}
(\boldsymbol{\Sigma}^{H}_{i}\mathbf{A}_{i,p}\boldsymbol{\Sigma}_{i}+\boldsymbol{\Upsilon}^{H}_{i}\tilde{\mathbf{A}}_{i,p}\boldsymbol{\Upsilon}_{i})^{-1} \times \\ (\boldsymbol{\Sigma}^{H}_{i}\mathbf{A}_{i,p}\mathbf{w}_{i,p}+
\boldsymbol{\Upsilon}^{H}_{i}\tilde{\mathbf{A}}_{i,p}\tilde{\mathbf{w}}_{i,p}) \ \ \text{for} \ \ 1<p<M \\
(\boldsymbol{\Sigma}^{H}_{i}\mathbf{A}_{i,p}\boldsymbol{\Sigma}_{i})^{-1}\boldsymbol{\Sigma}^{H}_{i}\mathbf{A}_{i,p}\mathbf{w}_{i,p}
 \ \ \ \  
  \text{for} \ \  p=1\\
 (\boldsymbol{\Upsilon}^{H}_{i}\tilde{\mathbf{A}}_{i,p}\boldsymbol{\Upsilon}_{i})^{-1}\boldsymbol{\Upsilon}^{H}_{i}\tilde{\mathbf{A}}_{i,p}\tilde{\mathbf{w}}_{i,p}
\ \ \
 \text{for} \ \ p=M
\end{array}\right.
 \end{small}
\end{equation}
 Therefore, $\boldsymbol{\theta}_{i,p}$ for $p\in \{1,\hdots,M\}$ are estimated in an iterative loop. With (\ref{Mstep_i}) and (\ref{theta_i_p_general}), it can be proven that the BCD algorithm leads to unique solutions and thus, convergence to at least a local maximizer, is ensured \cite{bertsekas1999nonlinear}. If the M-step is performed exactly (i.e., the BCD gives the exact minimizer of (25) and consequently, the M step is exactly solved), convergence of the EM algorithm to a stationary point is ensured (to avoid the unusual case of convergence to a saddle point, a proper initialization is required) \cite{mclachlan2007algorithm}, with a theoretical infinite number of iterations. Finally, in this case, convergence of the concentrated MLE is guaranteed for an infinite number of iterations since the value of the cost function at each step can either improve or maintain but cannot worsen \cite{vorobyov2005maximum}. In practice, only a finite number of iterations is  considered in each loop, so we might not attain local convergence.
However, we show in section  \ref{simu_sirp} the relatively good numerical stability of the algorithm.

 The scheme of the proposed algorithm is described in Algorithm 1. 

 \begin{algorithm}
\SetAlgorithmName{Algorithm 1}{}{} \caption{Relaxed concentrated ML based calibration algorithm}
\SetKwInOut{input}{input} \SetKwInOut{output}{output}
\SetKwInOut{initialize}{initialize}
\input{$D$, $M$, $B$, $\mathbf{C}_{i}$,  $\beta_{i}$, $\mathbf{x}$}
\output{$\hat{\boldsymbol{\theta}}$}
\initialize{$\hat{\boldsymbol{\Omega}}$ $\leftarrow$ $\boldsymbol{\Omega}_{\mathrm{init}}$, $\hat{\boldsymbol{\tau}}$ $\leftarrow$ $\boldsymbol{\tau}_{\mathrm{init}}$, $\hat{\boldsymbol{\theta}}\leftarrow\boldsymbol{\theta}_{\mathrm{init}}$}
\While
{stop criterion unreached}
{
\While 
{stop criterion unreached}
{
\ShowLn E-step: $\hat{\mathbf{w}}_{i}$ obtained from (\ref{wRealEstim}), $i \in \{1,\hdots, D\}$\\
\ShowLn M-step: $\hat{\boldsymbol{\theta}}_{i}$ obtained as follows, $i\in \{1, \hdots, D\}$ \\
\SetAlgoNoLine
\Indp \Indp 
\While   
{ stop criterion unreached}
{ $\hat{\boldsymbol{\theta}}_{i,p}$ obtained from (\ref{theta_i_p_general}), $p \in \{1, \hdots, M\}$
}
}
\ShowLn Obtain $\hat{\boldsymbol{\Omega}}$ from (\ref{OmegaEstim}) and (\ref{trace}),\\
\ShowLn Obtain $\hat{\boldsymbol{\tau}}$ from (\ref{tauExp})\\
}
\end{algorithm}

  \section{Structured Jones matrices}
  \label{structure}

  We recall that the output of Algorithm 1 is the estimate of each Jones matrix denoted by $\hat{\mathbf{J}}_{i,p}$ for $i\in \{1, \hdots, D\}$ and $p\in \{1, \hdots, M\}$. In the following, we consider the data model in (\ref{model_regime3}) for the specific 3DC calibration regime, and intend to estimate the unknown parameter vector of interest $\boldsymbol{\varepsilon}^{\mathrm{3DC}}$ in a sequential manner. To do so, we use an iterative estimation procedure by optimizing a cost function w.r.t.~one parameter while fixing the others.

    \textbf{\textit{1) Estimation of $\mathbf{g}_p$}}:
The diagonal elements of the gain matrix are given by     
      \begin{equation}
  \label{estim_gp}
 \hat{\mathbf{g}}_p= \argmin_{\mathbf{g}_p} \kappa(\mathbf{g}_p)
  \end{equation}
  where $\kappa(\mathbf{g}_p)=\sum_{i=1}^D  ||\hat{\mathbf{J}}_{i,p}-\mathbf{G}_p(\mathbf{g}_p)\mathbf{H}_{i,p}\mathbf{Z}_{i,p}\mathbf{F}_{i}||^2_{F}$. We rewrite the cost function as
  \begin{equation}
  \kappa(\mathbf{g}_p)=\sum_{i=1}^D\mathrm{Tr}\Big\{\Big(\hat{\mathbf{J}}_{i,p}-\mathbf{G}_p(\mathbf{g}_p)\mathbf{R}_{i,p}\Big)\Big(\hat{\mathbf{J}}_{i,p}-\mathbf{G}_p(\mathbf{g}_p)\mathbf{R}_{i,p}\Big)^H\Big\}
  \end{equation}
in which $\mathbf{R}_{i,p}=\mathbf{H}_{i,p}\mathbf{Z}_{i,p}\mathbf{F}_{i}$.
  The derivation of $ \kappa(\mathbf{g}_p)$ w.r.t. the \textit{k}-th element  $[\mathbf{g}_p]_k$ leads to
    \begin{equation}
    \label{derive_kappa}
  \frac{\partial \kappa(\mathbf{g}_p)}{\partial
[\mathbf{g}_p]_k}= \sum_{i=1}^D\mathrm{Tr}\Big\{-\mathbf{e}_k\mathbf{e}_k^T\mathbf{R}_{i,p}\hat{\mathbf{J}}_{i,p}^H+\mathbf{e}_k\mathbf{e}_k^T\mathbf{R}_{i,p}\mathbf{R}_{i,p}^H\mathbf{G}_p^H\Big\}.
    \end{equation}
    Let us denote $\mathbf{X}_{i,p}=\mathbf{R}_{i,p}\hat{\mathbf{J}}_{i,p}^H$ and $\mathbf{W}_{i,p}=\mathbf{R}_{i,p}\mathbf{R}_{i,p}^H$. From (\ref{derive_kappa}), we deduce the equation satisfied by the gain matrix
    \begin{align}
    \nonumber
\sum_{i=1}^D[\mathbf{X}_{i,p}]_{k,k} & =\sum_{i=1}^D[\mathbf{W}_{i,p}\hat{\mathbf{G}}_p^{H}]_{k,k} \\ & =
\sum_{i=1}^D[\mathbf{W}_{i,p}]_{k,k}[\hat{\mathbf{G}}_p^{\ast}]_{k,k}
\end{align}
    for $k \in \{1,2\}$, since $\mathbf{G}_p$ is a diagonal matrix. Therefore, each complex gain element is estimated as 
  \begin{equation}
  \label{estim_gp_final}
[\hat{\mathbf{g}}_p]_k=\Big(\sum_{i=1}^D[\mathbf{W}^{\ast}_{i,p}]_{k,k}\Big)^{-1} \sum_{i=1}^D[\mathbf{X}^{\ast}_{i,p}]_{k,k}.
\end{equation}

 \textbf{\textit{2) Estimation of $\boldsymbol{\alpha}_i$}}:
 We first need to estimate $\varphi_{i,p}$ (we recall that $\varphi_{i,p}=\eta_i u_p+\zeta_i v_p$). This is done as follows
       \begin{equation}
  \label{estim_phi_i_p}
 \hat{\varphi}_{i,p}= \argmin_{\varphi_{i,p}} \tilde{\kappa}(\varphi_{i,p})
  \end{equation}
  where  $\tilde{\kappa}(\varphi_{i,p})= ||\hat{\mathbf{J}}_{i,p}-\mathbf{G}_p\mathbf{H}_{i,p}\mathbf{Z}_{i,p}(\varphi_{i,p})\mathbf{F}_{i}||^2_{F}$. Taking the derivative of $\tilde{\kappa}(\varphi_{i,p})$ w.r.t. $\varphi_{i,p}$ and setting the result to zero, we obtain
 \begin{align}
 \nonumber
 & \mathrm{Tr}\Big\{j\exp^{-j\hat{\varphi}_{i,p}}\hat{\mathbf{J}}_{i,p}\mathbf{F}_{i}^H\mathbf{H}_{i,p}^H\mathbf{G}_p^H-j\exp^{j\hat{\varphi}_{i,p}}\mathbf{G}_p\mathbf{H}_{i,p}\mathbf{F}_{i}\hat{\mathbf{J}}_{i,p}^H\Big\} \\& =0
 \end{align}
 which leads to 
  \begin{equation}
  \exp\Big\{2j\hat{\varphi}_{i,p}\Big\}=\frac{\mathrm{Tr}\Big\{\mathbf{M}_{i,p}\Big\}}{\mathrm{Tr}\Big\{\mathbf{M}_{i,p}^H\Big\}}
 \end{equation}
 where $\mathbf{M}_{i,p}=\hat{\mathbf{J}}_{i,p}\mathbf{F}_{i}^H\mathbf{H}_{i,p}^H\mathbf{G}_p^H$.

In the case of a compact array, we can write for the \textit{i}-th source
   \begin{equation}
 \boldsymbol{\varphi}_i^T=\hat{\boldsymbol{\alpha}}_i^T
\boldsymbol{\Lambda}
  \end{equation}
 where $\boldsymbol{\varphi}_i=[\hat{\varphi}_{i,1},\hdots,\hat{\varphi}_{i,M}]^T$ and $\boldsymbol{\Lambda}= \begin{bmatrix}
   u_1, & \hdots & ,u_M\\
v_1, & \hdots &  ,v_M
  \end{bmatrix}$.
  Therefore, estimation of the directional shifts due to propagation in the ionosphere is given by
  \begin{equation}
  \label{estim_alpha_i}
 \hat{\boldsymbol{\alpha}}_i^T=
 \frac{\boldsymbol{\varphi}_i^T\boldsymbol{\Lambda}^H \begin{bmatrix}
  \sum_{p=1}^M v_p^2 & -\sum_{p=1}^M u_p v_p\\
-\sum_{p=1}^M v_p u_p &  \sum_{p=1}^M u_p^2
  \end{bmatrix}}{\sum_{p=1}^M u_p^2 \sum_{p=1}^M v_p^2 - (\sum_{p=1}^M u_p v_p)^2}.
 \end{equation}
 
   \textbf{\textit{3) Estimation of $\vartheta_i$}}:
  We consider the following minimization problem
  \begin{equation}
  \label{estim_thetai}
 \hat{\vartheta}_i= \argmin_{\vartheta_i} \sum_{p=1}^M ||\hat{\mathbf{J}}_{i,p}-\mathbf{G}_p\mathbf{H}_{i,p}\mathbf{Z}_{i,p}\mathbf{F}_{i}(\vartheta_{i})||^2_{F}.
  \end{equation}
  We assume a large number of antennas $M$ while the number of calibrator sources $D$ is relatively reduced, such that observations outnumber unknown parameters. For each source, the 1D optimization in (\ref{estim_thetai}) can be computed in a reasonable computational time through a classical data grid search or a Newton type algorithm.

 Finally, the proposed algorithm for the structured Jones matrices case regarding 3DC calibration regime is given in Algorithm 2.
 \begin{algorithm}
\SetAlgorithmName{Algorithm 2}{}{} \caption{Case of structured Jones matrices}
\SetKwInOut{input}{input} \SetKwInOut{output}{output}
\SetKwInOut{initialize}{initialize}
\input{$D$, $M$, $B$, $\mathbf{C}_{i}$,  $\beta_{i}$, $\mathbf{x}$, $\hat{\mathbf{J}}_{i,p}$ as output of Algorithm 1, $i \in \{1, \hdots, D\}$ and $p \in \{1, \hdots, M\}$} 
\output{$\hat{\boldsymbol{\varepsilon}}^{\mathrm{3DC}}$}
\initialize{$\hat{\boldsymbol{\varepsilon}}^{\mathrm{3DC}} \leftarrow \boldsymbol{\varepsilon}^{\mathrm{3DC}}_{\mathrm{init}}$}
\While
{stop criterion unreached}
{\setcounter{AlgoLine}{0} \ShowLn Obtain $\hat{\vartheta_i}$ from (\ref{estim_thetai}), $i \in \{1, \hdots, D\}$ \\
\ShowLn Obtain $\hat{\mathbf{g}}_p$ from (\ref{estim_gp_final}), $p \in \{1, \hdots, M\}$\\
\ShowLn Obtain $\boldsymbol{\alpha}_i=[\eta_i,\zeta_i]^T$ from (\ref{estim_alpha_i}), $i \in \{1, \hdots, D\}$ \\
}
\end{algorithm}

  \section{Numerical simulations}
  \label{simus}
  
  In this part, we first aim to assess the statistical performance of Algorithm 1, when the noise model matches our noise assumption, i.e., a SIRP noise modeling. Afterwards, we intend to study our proposed algorithm, in a more realistic scenario where outliers are present. More specifically, the sky is composed of $D$ known bright calibrator sources but also $D'$ weak non-calibrator sources which are absorbed in the noise component and act as outliers. Under such assumption, we compare our scheme with the recently introduced robust calibration approach based on Student's t \cite{yatawatta2014robust} and with the traditional Gaussian cases \cite{yatawatta2009radio}. Finally, we apply Algorithm 2 to the introduced 3DC calibration regime where Jones matrices are structured.
  
  \subsection{Numerical results under SIRP noise}
  \label{simu_sirp}
  
The unknown parameters to estimate, $\boldsymbol{\theta}$ given in (\ref{express_theta}), correspond to the real and imaginary parts of the entries of all Jones matrices. The additive noise in (\ref{express_x}) is assumed to follow a SIRP as in (\ref{sirp}) and the number of Monte Carlo runs is set to 100. 

 \begin{figure}[t!]
 \centering
\subfigure[]{\label{fig:SIRV-aa}\includegraphics[width=0.45\textwidth,height=0.35\textwidth]{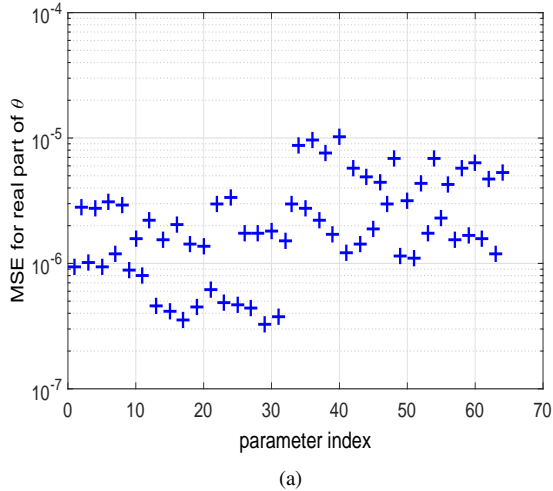}} \hspace{0.5pt} 
\subfigure[]{\label{fig:SIRV-bb}\includegraphics[width=0.45\textwidth,height=0.35\textwidth]{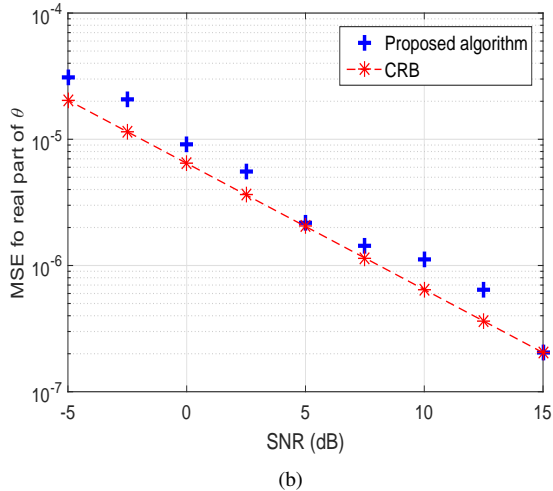}} 
\caption{\label{fig1:SIRV}(a) MSE of the real part of the first $32$ unknown parameters for a given SNR, (b) MSE vs. SNR for the real part of a given unknown parameter and the corresponding CRB, for $D=2$ bright signal sources and $M=8$ antennas, leading to $128$ real unknown parameters of interest to estimate and $224$ measurements. }
\end{figure}

\begin{figure}[t!]
 \centering
\subfigure[]{\label{fig2:SIRV-Conv}\includegraphics[width=0.45\textwidth,height=0.35\textwidth]{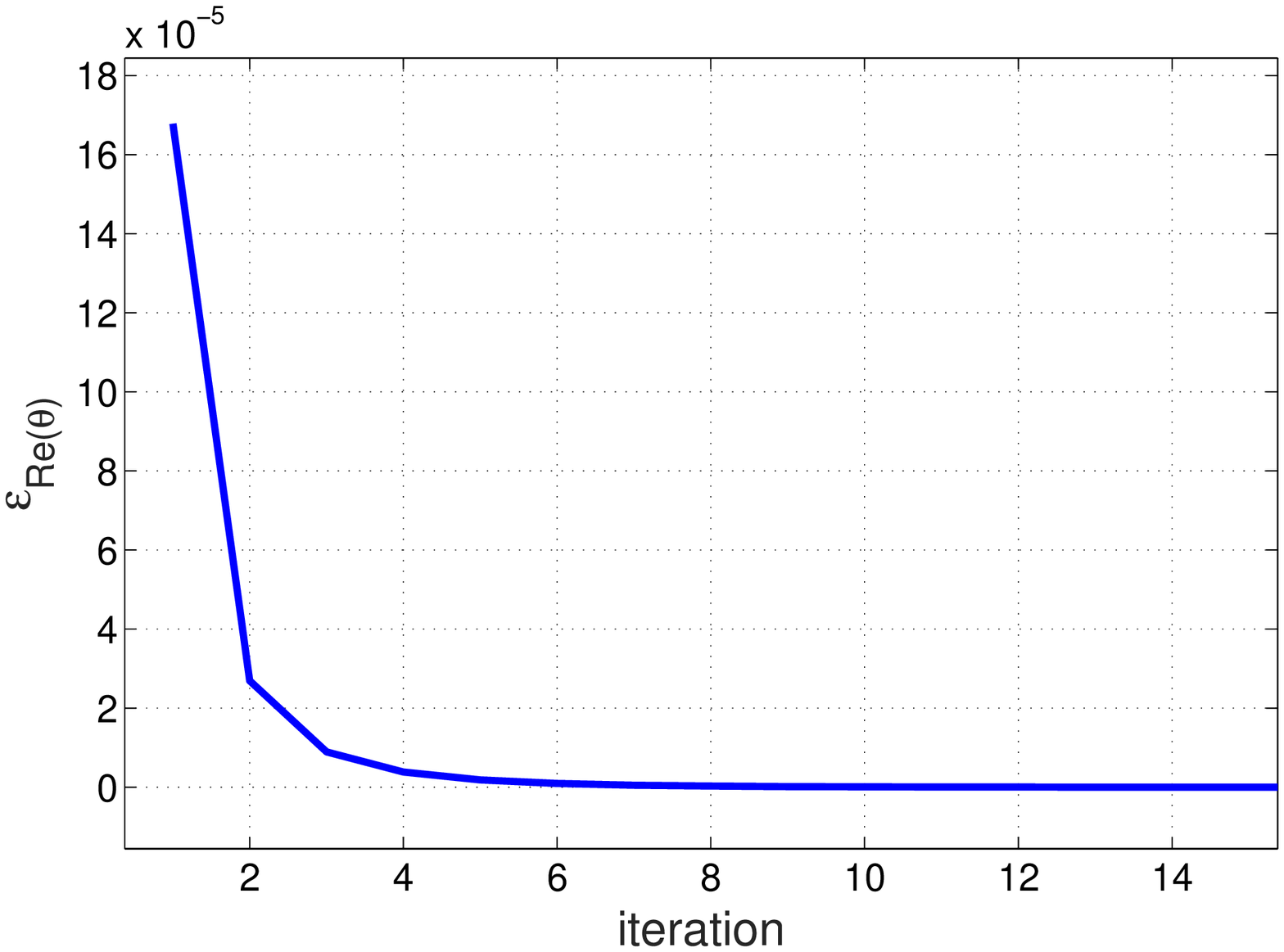}} \hspace{0.5pt} 
\subfigure[]{\label{fig2:SIRV-Conv_log}\includegraphics[width=0.45\textwidth,height=0.35\textwidth]{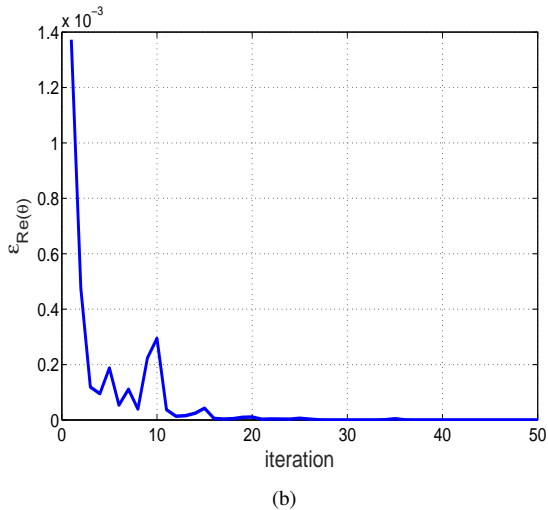}} 
\caption{\label{fig2:SIRV-Conv}$\epsilon_{\Re\left\{\boldsymbol{\theta}\right\}}^h$ as function of the \textit{h}-th iteration, for loop in line 2 (a), and  in line 1 (b) from Algorithm 1. }
\end{figure}

In order to evaluate the estimation performance of the relaxed concentrated ML based calibration algorithm, we fix the noise distribution, e.g., as a Student's t and make use of the Cram\'{e}r-Rao bound (CRB) \cite{stoica2005spectral}. To do so, each random texture component is supposed to follow an inverse gamma distribution\cite{balleri2007maximum}.
As an example,
\begin{equation}
\tau_{pq} \sim \mathcal{IG}(\nu/2,\nu/2) ,
\end{equation}
with $\nu$ degrees of freedom \cite{ollila2012complex} and we choose for example $[\boldsymbol{\Omega}]_{k,l}=\sigma^2 0.9^{|k-l|}\exp^{j\frac{\pi}{2}(k-l)}$.

The covariance inequality principle states that, under quite general/weak conditions, the variance satisfies
  \begin{equation}
\mathrm{MSE}([\hat{\boldsymbol{\theta}}]_k)=\mathrm{E}\Big\{\Big([\hat{\boldsymbol{\theta}}]_k-[\boldsymbol{\theta}]_k\Big)^2\Big\}\geqslant[\mathrm{CRB}(\boldsymbol{\theta})]_{k,k}
\end{equation}
where the CRB is given as the inverse of the Fisher information matrix (FIM) $\mathbf{F}$. A Slepian-Bangs type formula of the FIM for SIRP observations is given in \cite{besson2013fisher}  which can be adapted to our case and reads 
\begin{equation}
[\mathbf{F}]_{k,l}= 2\frac{\nu+4}{\nu+5}\sum_{pq}\Re\left\{
\frac{\partial \tilde{\mathbf{v}}^H_{pq}(\boldsymbol{\theta})}{\partial[\boldsymbol{\theta}]_k}\boldsymbol{\Omega}^{-1}\frac{\partial \tilde{\mathbf{v}}_{pq}(\boldsymbol{\theta})}{\partial[\boldsymbol{\theta}]_l}\right\}.
\end{equation}
Note that the noise parameters are decoupled from the parameters of interest. Consequently, only the part corresponding to the latter is kept in the FIM expression.

 First, in Fig. \ref{fig:SIRV-aa}, we plot the mean square error (MSE) of the real part of each unknown parameter, obtained with Algorithm 1, for a signal-to-noise ratio $\mathrm{SNR}=15 \mathrm{dB}$. We only plot the parameters relative to one given source, the behavior being the same for any source.  
 We also compare the MSE of one given parameter, as a function of the SNR, to its corresponding CRB  in Fig. \ref{fig:SIRV-bb}. This enables to assess the statistical performance of Algorithm 1, and we notice that the MSE approaches the CRB. The small gap between the bound and the algorithm is explained by the relaxed nature hypothesis used in the design of Algorithm 1. Indeed, we have assumed unknown and deterministic texture parameters when we derived the estimates using Algorithm 1, but in the data model, these parameters are generated as random variables following inverse gamma distribution and the CRB was derived using the prior of the pdf of the texture.

Second, we now aim to investigate numerically the convergence properties of our algorithm, which is composed of 3 loops. In each of these loops, $\boldsymbol{\theta}$ is updated at each iteration. We therefore consider the following quantity
\begin{equation}
\epsilon_{\Re\left\{\boldsymbol{\theta}\right\}}^h=||\Re\left\{\boldsymbol{\theta}^h -\boldsymbol{\theta}^{h-1}\right\}||^2_2
\end{equation}
where $h$ refers to the \textit{h}-th iteration.
In Fig. \ref{fig2:SIRV-Conv} we present the convergence rate of loops described in Algorithm 1 at line 1 and 2 (the analysis of convergence of the third loop, given in line 5, has the same behavior as the loop in line 2 and thus, is not reported here).
We note that around 5 iterations are required in loop 2 to attain convergence while approximately 20 iterations are needed for the algorithm to be stable, in loop 1. Nevetheless, in simulations, we notice that only 3 to 4 iterations are sufficient to get close to the CRB.

\begin{figure}[t!]
 \centering
\subfigure[]{\label{fig:outlier_idem_temps-aa}\includegraphics[width=0.45\textwidth,height=0.35\textwidth]{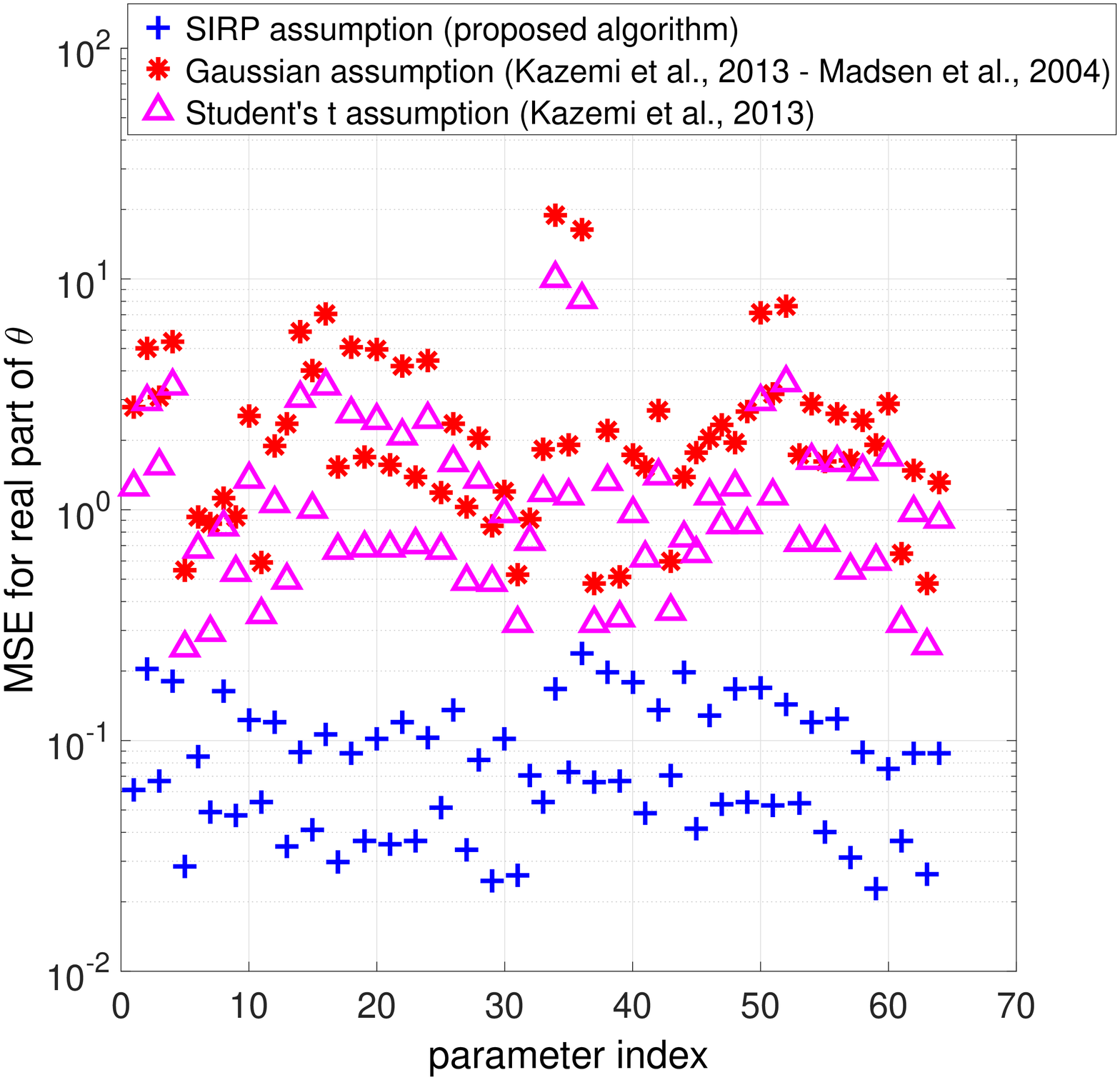}} \hspace{0.5pt} 
\subfigure[]{\label{fig:outlier_idem_temps-bb}\includegraphics[width=0.45\textwidth,height=0.35\textwidth]{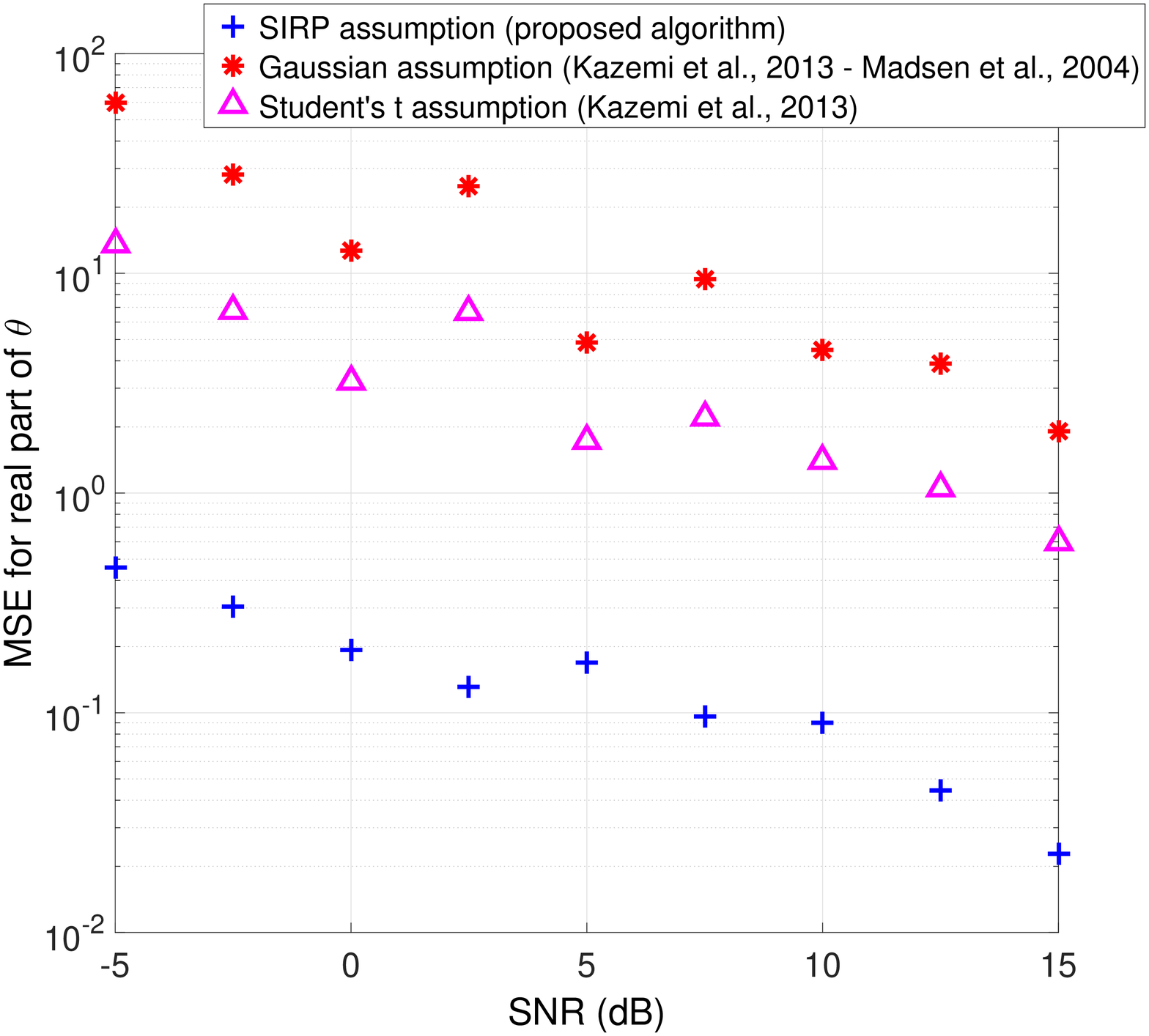}} 
\caption{\label{fig1:outlier_idem_temps}(a) MSE of the real part of the $64$ unknown parameters for a given SNR, (b) MSE vs. SNR for the real part of a given unknown parameter, for $D=2$, $M=8$ and $D'=8$, leading to $128$ real parameters of interest to estimate and $224$ measurements. }
\end{figure}

\begin{figure}[t!]
 \centering
\subfigure[]{\label{fig:gain_param_AjoutNabil}\includegraphics[width=0.45\textwidth,height=0.35\textwidth]{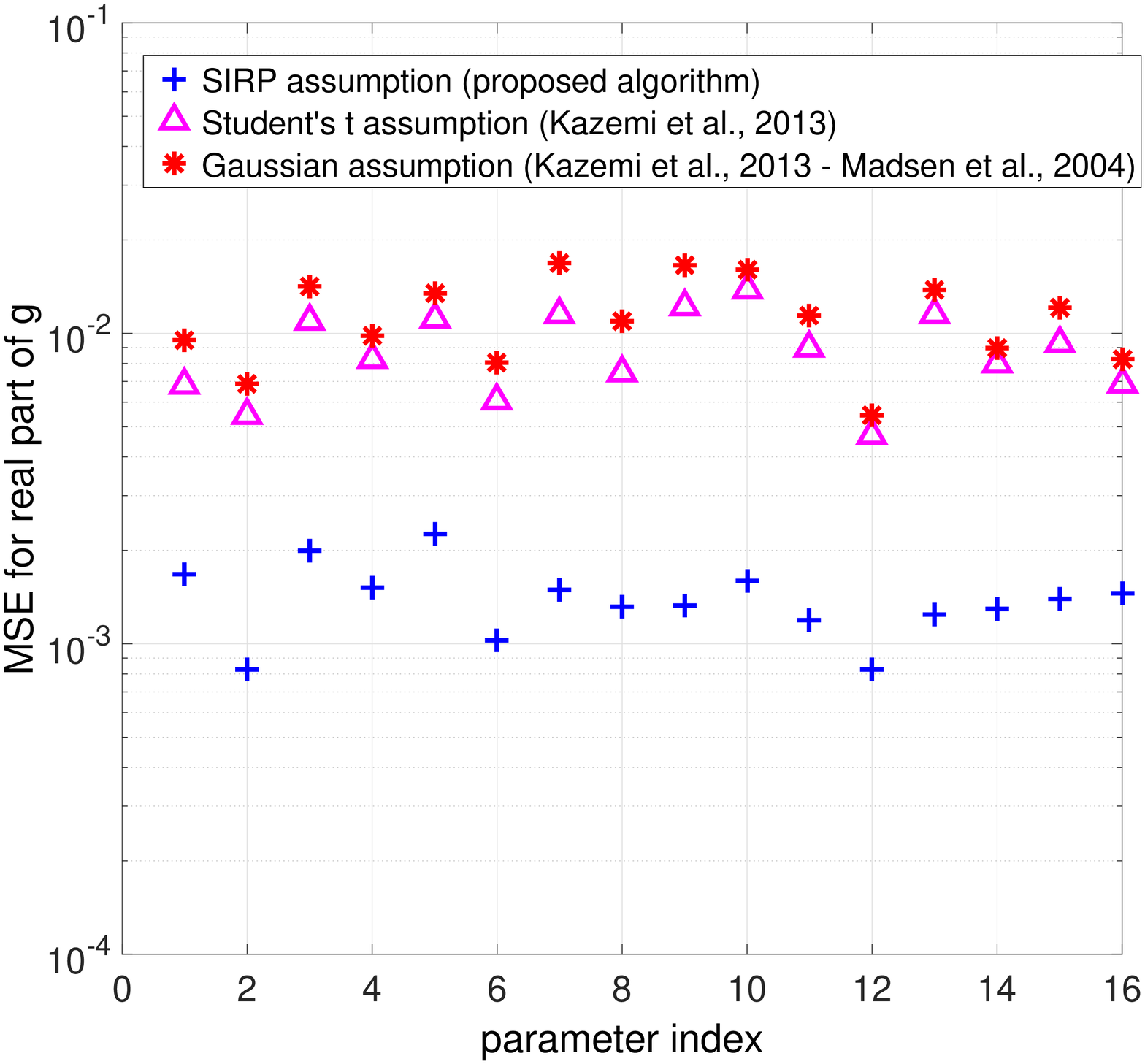}} \hspace{0.5pt} 
\subfigure[]{\label{fig:ionos_param}\includegraphics[width=0.45\textwidth,height=0.35\textwidth]{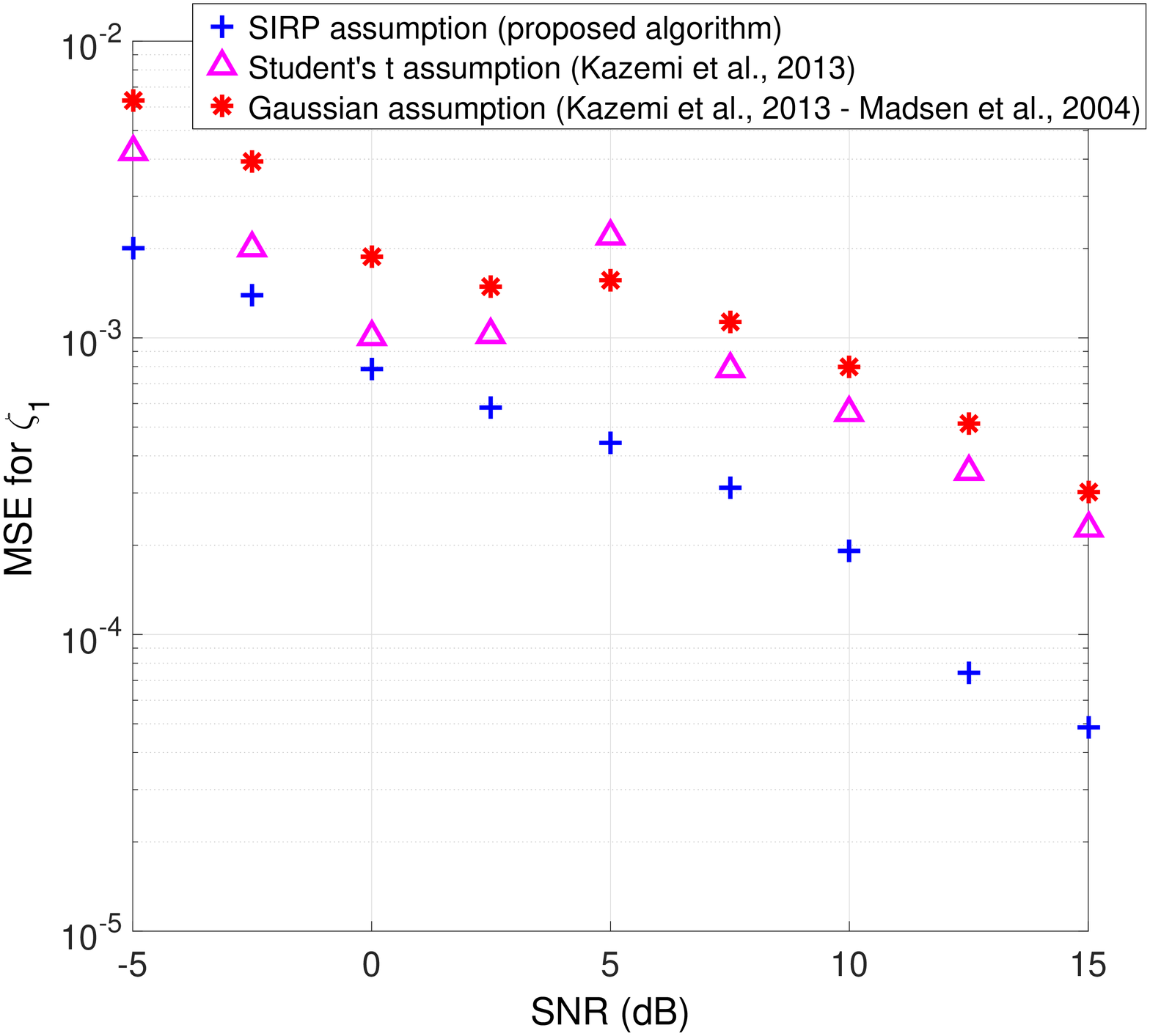}} 
\caption{\label{fig:all_structured}(a) MSE of the real part of the $16$ complex gains for a given SNR, (b) MSE of $\zeta_1$ vs. SNR, for $D=2$, $M=8$, and $D'=4$, leading to $38$ real parameters of interest to estimate and $224$ measurements.}
\end{figure}

\subsection{Numerical results using a realistic model}

Let us investigate the robustness of our proposed calibration procedure in a realistic situation, and compare it with the state-of-the-art. To do so, we consider $D$ calibrator sources, $D'$ weak outlier sources and Gaussian background noise in our data model.
The real parameters of interest to estimate still correspond to the real and imaginary entries of the Jones matrices associated to the calibrator sources paths. In the following, we compare Algorithm 1 with i) the calibration approach exposed in \cite{yatawatta2014robust} which assumes a Student's t noise modeling using the so-called expectation-conditional maximization either algorithm \cite{liu1995ml,li2006t} and ii) the traditional calibration scheme based on zero-mean white Gaussian noise modeling using a least squares approach \cite{yatawatta2009radio}. The comparative results are plotted in Fig. \ref{fig1:outlier_idem_temps}, for the same computation times. We notice better estimation performance with Algorithm 1, since we did not specify any particular noise distribution in our procedure and a SIRP includes many different types of distributions. Furthermore, no assumption was made about independent entries in the noise vector, thus ensuring more flexibility and robustness.

\subsection{Structured case}

In order to compare our proposed global algorithm (Algorithm 1 followed by Algorithm 2) with the approach based on Student's t  \cite{yatawatta2014robust} and the Gaussian case \cite{yatawatta2009radio} which were both introduced in the non-structured case, we apply Algorithm 2 on the output of these two latter algorithms.

Each Jones matrix is generated according to (\ref{model_regime3}) with $\mathbf{g}=[\mathbf{g}^T_1,\hdots,\mathbf{g}^T_M]^T$. In all exposed simulations, we consider similar computation times for the three presented patterns. We show the results only for the complex gains, cf. Fig. \ref{fig:gain_param_AjoutNabil}, and the source offset $\zeta_1$, cf. Fig \ref{fig:ionos_param}, due to lack of space, the behavior being the same for the other parameters, i.e., $\vartheta_1$, $\vartheta_2$, $\eta_2$, $\eta_1$ and $\zeta_2$. In the case of structured Jones matrices, adapted to the 3DC calibration regime, and in the presence of outliers, we still notice the better performances of Algorithm 1, compared to the state-of-the-art. This is expected since better estimation of the Jones entries leads to better estimation of the physical parameters describing the structured Jones matrices.

 \section{Conclusion} 
 \label{ccl}
 
 In this paper, we proposed a robust calibration technique where perturbation effects are modeled thanks to Jones matrices. To deal with the presence of outliers in our data, the introduced ML estimation method is based on SIRP noise modeling, leading to a relaxed concentrated ML based calibration algorithm. Numerical simulations show that the proposed algorithm is more robust to the presence of outliers in comparison with the state-of-the-art, for both non-structured and structured Jones matrices, with a reasonable computational complexity.

\appendices

\section{}

We describe here the corresponding expressions of (\ref{tauExp}) and (\ref{OmegaEstim}) when we assume a different $\boldsymbol{\Omega}_{pq}$ for $p<q, \ \ p,q \in \{1, \ldots, M\}$. In this case, the log-likelihood function is written as
\begin{align}
\nonumber
 & \log f(\mathbf{x}|\boldsymbol{\theta },
\boldsymbol{\tau}, \boldsymbol{\Omega}_{12},\boldsymbol{\Omega}_{13},\hdots,\boldsymbol{\Omega}_{(M-1)M}) =
-4B\log\pi \\  &
-4\sum_{pq}\log\tau_{pq}-
\sum_{pq}\log |\boldsymbol{\Omega}_{pq}|-\sum_{pq}\frac{1}{\tau_{pq}}\mathbf{a}_{pq}^{H}(\boldsymbol{\theta})
\boldsymbol{\Omega}_{pq}^{-1}\mathbf{a}_{pq}(\boldsymbol{\theta}).
\end{align}
For each antenna pair, the texture estimate is given by
\begin{equation}
 \hat{\tau}_{pq} = \frac{1}{4} \mathbf{a}_{pq}^{H}(\boldsymbol{\theta})
\boldsymbol{\Omega}_{pq}^{-1} \mathbf{a}_{pq}(\boldsymbol{\theta})
\end{equation}
while the speckle covariance estimate reads
\begin{equation}
\label{new_Omega}
 \hat{\boldsymbol{\Omega}}_{pq}^{h+1} = 4
 \frac{\mathbf{a}_{pq}(\boldsymbol{\theta})
\mathbf{a}_{pq}^{H}(\boldsymbol{\theta})}{\mathbf{a}_{pq}^{H}(\boldsymbol{\theta})\Big(\hat{\boldsymbol{\Omega}}_{pq}^{h}\Big)^{-1}\mathbf{a}_{pq}(\boldsymbol{\theta})}.
\end{equation}
The remainder of the algorithm is straightforwardly obtained using (\ref{new_Omega}).

\section{}
We present here the steps to obtain (\ref{theta_i_p_general}).
Firstly, for sake of clarity, let us denote $\mathbf{c}_{i}=[ c_{i_{1}},c_{i_{2}},c_{i_{3}},c_{i_{4}}]^T$ to refer to the four entries of the vectorization of source coherency matrix $\mathbf{C}_{i}$. Likewise, for the \textit{i}-th source, we write $
\mathbf{J}_{i,p}(\boldsymbol{\theta}_{i,p}) = \begin{bmatrix}
    p_{i_{1}} &  p_{i_{2}}\\
    p_{i_{3}} & p_{i_{4}}
  \end{bmatrix}$ for the \textit{p}-th antenna and $ 
\mathbf{J}_{i,q}(\boldsymbol{\theta}_{i,q}) = \begin{bmatrix}
    q_{i_{1}} &  q_{i_{2}}\\
    q_{i_{3}} & q_{i_{4}}
  \end{bmatrix}$ for the \textit{q}-th antenna, i.e., $\boldsymbol{\theta}_{i,p}=[p_{i_{1}},p_{i_{2}},p_{i_{3}},p_{i_{4}}]^{T}$ and $\boldsymbol{\theta}_{i,q}=[q_{i_{1}},q_{i_{2}},q_{i_{3}},q_{i_{4}}]^{T}$. 
Using these latter notation, we obtain (\ref{form_pq})
where 
\begin{equation}
\boldsymbol{\Sigma}_{i,q}= \begin{bmatrix}
   \alpha_{i,q} & \beta_{i,q} & 0 & 0\\
0 & 0 & \alpha_{i,q} & \beta_{i,q}\\
     \gamma_{i,q} & \rho_{i,q} & 0 & 0\\
0 & 0 & \gamma_{i,q} & \rho_{i,q}
  \end{bmatrix}
  \end{equation}
   in which $\alpha_{i,q}=q^{\ast}_{i_{1}}c_{i_{1}}+q^{\ast}_{i_{2}}c_{i_{3}}$, $\beta_{i,q}=q^{\ast}_{i_{1}}c_{i_{2}}
    + q^{\ast}_{i_{2}}c_{i_{4}}$, $\gamma_{i,q}=q^{\ast}_{i_{3}}c_{i_{1}}+q^{\ast}_{i_{4}}c_{i_{3}}$ and $\rho_{i,q}=q^{\ast}_{i_{3}}c_{i_{2}}
    + q^{\ast}_{i_{4}}c_{i_{4}}$.\\
We also obtain (\ref{form_qp})
where
\begin{equation}
   \boldsymbol{\Upsilon}_{i,q}=\begin{bmatrix}
   \lambda_{i,q} & \mu_{i,q} & 0 & 0\\
\nu_{i,q} & \xi_{i,q} & 0 & 0 \\
      0 & 0 & \lambda_{i,q} & \mu_{i,q} \\
0 & 0 & \nu_{i,q} & \xi_{i,q}
  \end{bmatrix}
  \end{equation}
   in which $\lambda_{i,q}=q_{i_{1}}c_{i_{1}}+q_{i_{2}}c_{i_{2}}$, $\mu_{i,q}=q_{i_{1}}c_{i_{3}}
    + q_{i_{2}}c_{i_{4}}$, $\nu_{i,q}=q_{i_{3}}c_{i_{1}}+q_{i_{4}}c_{i_{2}}$ and $\xi_{i,q}=q_{i_{3}}c_{i_{3}}
    + q_{i_{4}}c_{i_{4}}$.

Finally, the cost function in (\ref{Mstep_i_p}) can be written as 
\begin{align}
\label{compact_exp}
& \nonumber \phi_{i}(\boldsymbol{\theta}_{i,p}) = \Big(\mathbf{w}_{i,p} - \mathbf{u}_{i,p}(\boldsymbol{\theta}_{i,p})\Big)^H\mathbf{A}_{i,p}\Big(\mathbf{w}_{i,p} - \mathbf{u}_{i,p}(\boldsymbol{\theta}_{i,p})\Big) + 
\\ &
\Big(\tilde{\mathbf{w}}_{i,p} - \tilde{\mathbf{u}}_{i,p}(\boldsymbol{\theta}_{i,p})\Big)^H\tilde{\mathbf{A}}_{i,p}\Big(\tilde{\mathbf{w}}_{i,p} - \tilde{\mathbf{u}}_{i,p}(\boldsymbol{\theta}_{i,p})\Big)+\mathrm{Constant}
\end{align}
where $\mathbf{w}_{i,p}=[\mathbf{w}_{i,p(p+1)}^T, \hdots,\mathbf{w}_{i,p M}^T]^T$, $\mathbf{u}_{i,p}(\boldsymbol{\theta}_{i,p})=[  \mathbf{u}_{i,p(p+1)}^T(\boldsymbol{\theta}_{i,p}),\hdots,\mathbf{u}_{i,p M}^T(\boldsymbol{\theta}_{i,p})]^T$ and $\mathbf{A}_{i,p}=\mathrm{bdiag}\{\beta_{i}\tau_{p(p+1)} \boldsymbol{\Omega},\hdots, \beta_{i}\tau_{pM}\boldsymbol{\Omega}\}^{-1}$.

 Furthermore, we have $\tilde{\mathbf{w}}_{i,p}=[
\mathbf{w}^{\ast^T}_{i,1p},\hdots,\mathbf{w}^{\ast^T}_{i,(p-1)p}]^T$,  $\tilde{\mathbf{u}}_{i,p}(\boldsymbol{\theta}_{i,p})=[
\mathbf{u}^{\ast^T}_{i,1p}(\boldsymbol{\theta}_{i,p}),\hdots,\mathbf{u}^{\ast^T}_{i,(p-1)p}(\boldsymbol{\theta}_{i,p})]^T$ and $\tilde{\mathbf{A}}_{i,p}=
\mathrm{bdiag}\{\beta_{i}\tau_{1p} \boldsymbol{\Omega}^{\ast}, \hdots, \beta_{i}\tau_{(p-1)p}\boldsymbol{\Omega}^{\ast}\}^{-1}$.

We make use of (\ref{form_pq}) in what follows
\begin{equation}
\label{express_delta}
\mathbf{u}_{i,p}(\boldsymbol{\theta}_{i,p})=\begin{bmatrix}
   \mathbf{u}_{i,p(p+1)}(\boldsymbol{\theta}_{i,p})\\
\vdots\\
\mathbf{u}_{i,p M}(\boldsymbol{\theta}_{i,p})
  \end{bmatrix} = \begin{bmatrix}
  \boldsymbol{\Sigma}_{i,p+1}\boldsymbol{\theta}_{i,p}\\
\vdots\\
\boldsymbol{\Sigma}_{i,M}\boldsymbol{\theta}_{i,p}
  \end{bmatrix} = \boldsymbol{\Sigma}_{i} \boldsymbol{\theta}_{i,p}
\end{equation}
where $\boldsymbol{\Sigma}_{i}=[ \boldsymbol{\Sigma}_{i,p+1}^T,\cdots,\boldsymbol{\Sigma}_{i,M}^T]^T$.
Likewise, we use (\ref{form_qp}) in 
\begin{equation}
\label{express_upsilon}
\tilde{\mathbf{u}}_{i,p}(\boldsymbol{\theta}_{i,p})=\begin{bmatrix}
   \mathbf{u}^{\ast}_{i,1 p}(\boldsymbol{\theta}_{i,p})\\
\vdots\\
\mathbf{u}^{\ast}_{i,(p-1) p}(\boldsymbol{\theta}_{i,p})
  \end{bmatrix} = \begin{bmatrix}
  \boldsymbol{\Upsilon}^{\ast}_{i,1}\boldsymbol{\theta}_{i,p}\\
\vdots\\
\boldsymbol{\Upsilon}^{\ast}_{i,p-1}\boldsymbol{\theta}_{i,p}
  \end{bmatrix} = \boldsymbol{\Upsilon}_{i}\boldsymbol{\theta}_{i,p}
\end{equation}
in which $\boldsymbol{\Upsilon}_{i}=[ \boldsymbol{\Upsilon}^{\ast^T}_{i,1},\cdots,\boldsymbol{\Upsilon}^{\ast^T}_{i,p-1}]^T$.
Inserting (\ref{express_delta}) and (\ref{express_upsilon}) into (\ref{compact_exp}) and taking the derivative w.r.t. $\boldsymbol{\theta}_{i,p}$ leads to the expressions in (\ref{theta_i_p_general}), using the fact that $\mathbf{A}_{i,p}$ and $\tilde{\mathbf{A}}_{i,p}$ are Hermitian.

 \bibliographystyle{IEEEtran}
\bibliography{biblio}

\begin{IEEEbiography}
[{\includegraphics[width=1in,height=1.25in,clip,keepaspectratio]{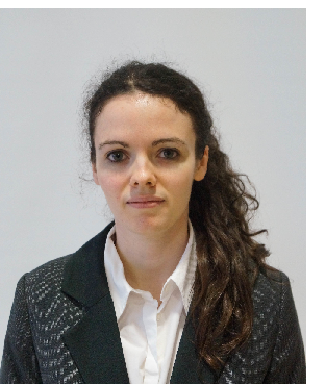}}]
{Virginie Ollier} was born in Beaumont, France, in 1991. She graduated from Ecole Centrale Marseille and received a Master Research degree in Signal and Image Processing from Ecole Centrale Marseille in 2015. She is currently pursuing a Ph.D. degree in signal processing in University of Paris Saclay and is a member of  SATIE laboratory (Syst\`{e}mes et Applications des Technologies de l'Information et de l'Energie) and L2S laboratory (Laboratoire des Signaux et Syst\`{e}mes).
Her research interests include estimation theory and statistical signal processing, especially for applications in robust array signal processing.
\end{IEEEbiography}

\begin{IEEEbiography}
[{\includegraphics[width=1in,height=1.25in,clip,keepaspectratio]{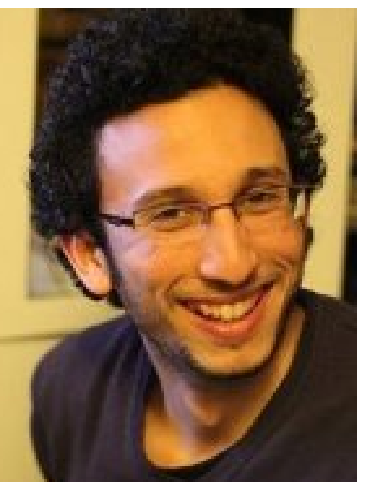}}]
{Mohammed Nabil El Korso}was born in Oran, Algeria. He received the M.Sc. in Electrical Engineering from the National Polytechnic School, Algeria in 2007. He obtained the Master Research degree in Signal and Image Processing from Paris-Sud XI University, France in 2008. In 2011, he obtained his Ph.D. degree from Paris-Sud XI University. From 2011 to 2012, he was a research scientist in the Communication Systems Group at Technische Universit\"{a}t Darmstadt, Germany. He was Assistant Professor at Ecole Normale Sup\'{e}rieure de Cachan from 2012 to 2013. Currently, he is Assistant Professor at University of Paris Ouest Nanterre la Defense and a member of LEME (EA4416) laboratory. His research interests include statistical signal processing, estimation/detection theory with applications to array signal processing. 
\end{IEEEbiography}

\begin{IEEEbiography}
[{\includegraphics[width=1in,height=1.25in,clip,keepaspectratio]{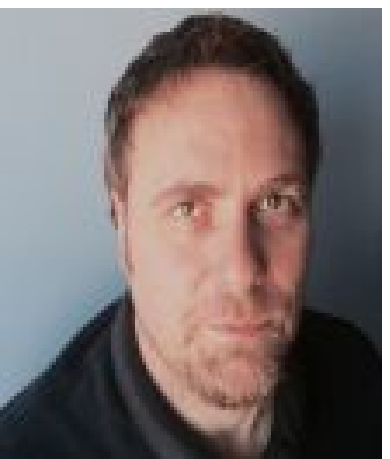}}]
{R\'{e}my Boyer} received the M.Sc. and Ph.D degrees from the Ecole Nationale Sup\'{e}rieure des Telecommunications (ENST-Paris or T\'{e}l\'{e}com ParisTech) in 1999 and 2002, respectively, in statistical signal processing. From 2002 to 2003, he was a postdoctoral fellow during six months at Sherbrooke University (Canada).  From 2003, he is an associate professor at University of
Paris-Sud - Laboratory of Signals and Systems (L2S). From 2011 to 2012, R\'{e}my Boyer was a visiting researcher at the SATIE Laboratory (Ecole Normale Sup\'{e}rieure de Cachan) and at the University of Aalborg (Danemark). His teaching activities include basic and advanced notions of statistical signal processing, in particular in the context of the Master2R ATSI ("Automatique et Traitement du Signal et des Images"). R\'{e}my Boyer received an "Habilitation \`{a} Diriger des Recherches (HDR)" from University of
Paris-Sud in December 2012. His research interests include compressive sampling, array signal processing, Bayesian performance bounds for parameter estimation and detection, security in mobile networks as well as numerical linear and multi-linear algebra. 
\end{IEEEbiography}

\begin{IEEEbiography}
[{\includegraphics[width=1in,height=1.25in,clip,keepaspectratio]{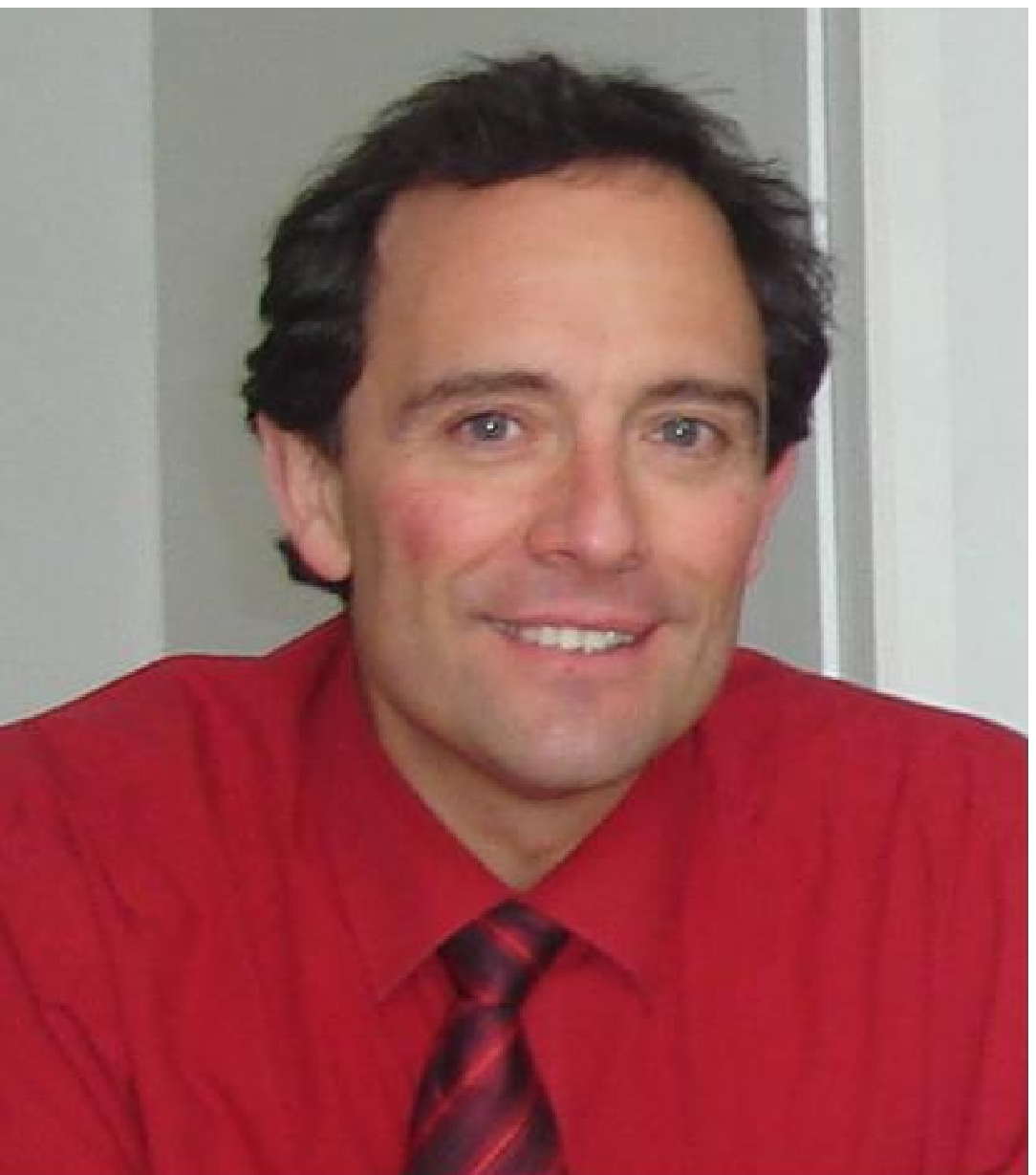}}]
{Pascal Larzabal} (M'93) was born in the Basque country in the south of France in 1962. He entered the Ecole Normale Sup\'{e}rieure of Cachan (France), in 1985 where he received the Agr\'{e}gation in Electrical Engineering in 1988. He received the PhD in 1992 and the "Habilitation \`{a} Diriger des Recherches" in 1998. He is now Professor of Electrical Engineering at University of ParisSud 11, France. He teaches electronic, signal processing,  control and mathematics. From 1998 to 2003 he was the director of Electrical Engineering IUP of  the University Paris-Sud. From March 2003 to March 2007 he was at the head of electrical engineering department in the Institut Universitaire de Technologie of Cachan. Since January 2007 he is the director of the laboratory SATIE in Paris. His research concerns estimation in array processing and spectral analysis for wavefront identification, radars, communications, tomography and medical imaging. His recent works concern geographical positioning and radio astronomy.

\end{IEEEbiography}

\begin{IEEEbiography}
[{\includegraphics[width=1in,height=1.25in,clip,keepaspectratio]{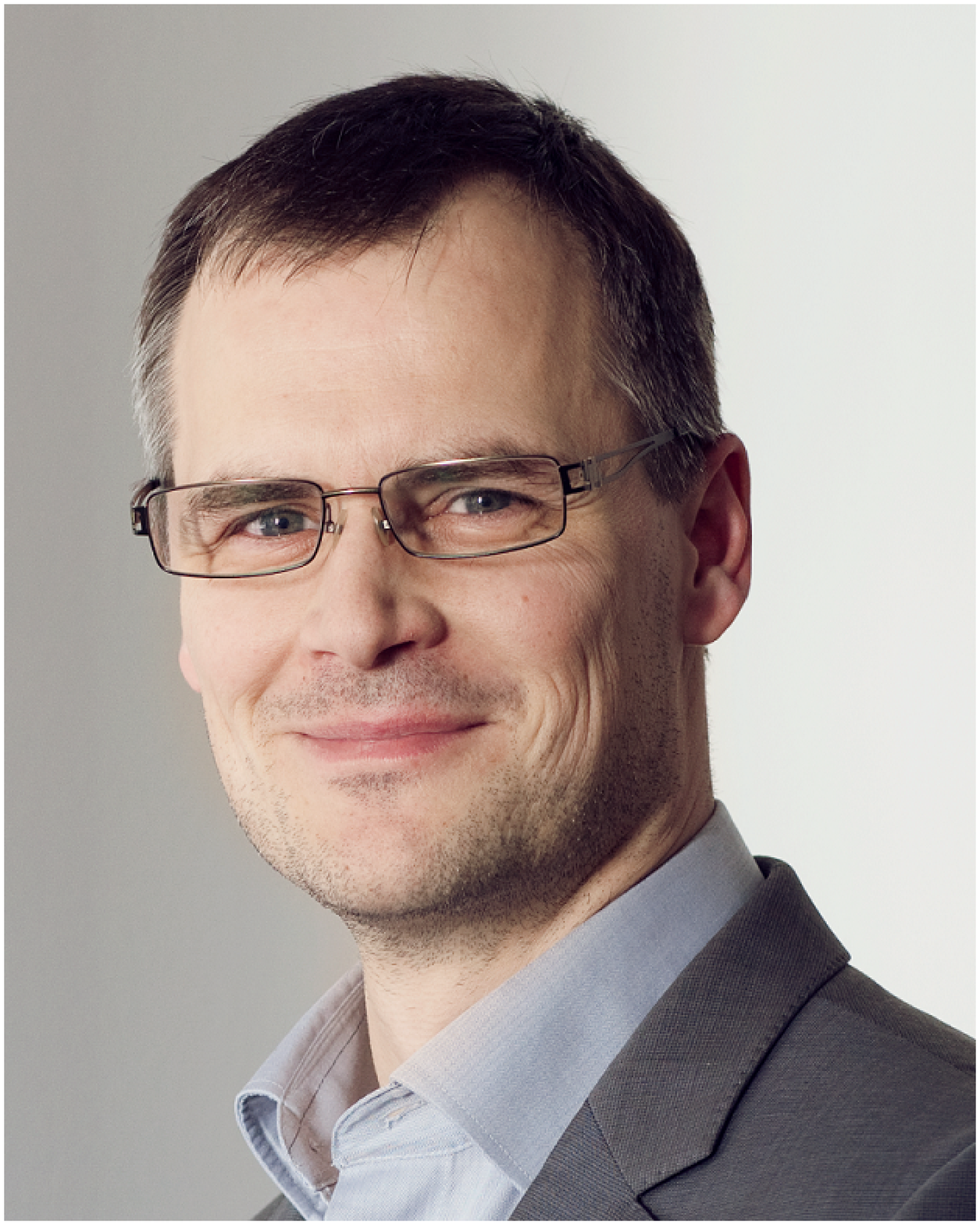}}]
{Marius Pesavento} (M'00) received the Dipl.-Ing. and M.Eng. degrees from Ruhr-University Bochum, Bochum, Germany, and McMaster University,
Hamilton, ON, Canada, in 1999 and 2000, respectively, and the Dr.-Ing. degree in electrical engineering from Ruhr-University Bochum in 2005.
Between 2005 and 2009, he held research positions in two start-up companies. In 2010, he became an Assistant Professor for Robust Signal Processing and a Full Professor for Communication Systems in 2013, at the Department of Electrical Engineering and Information Technology, Technical University Darmstadt, Darmstadt, Germany. His research interests include robust signal processing and adaptive beamforming, high-resolution sensor array processing, multiantenna and multiuser communication systems, distributed, sparse, and mixed-integer optimization techniques for signal processing and communications, statistical signal processing, spectral analysis, and parameter estimation. He has received the 2003 ITG/VDE Best Paper Award, the 2005 Young Author Best Paper Award of the IEEE Transactions on Signal Processing, and the 2010 Best Paper Award of the CrownCOM conference. He is a Member of the Editorial Board of the EURASIP Signal Processing Journal, and served as an Associate Editor for the IEEE Transactions on Signal Processing in 2012-2016. He is a Member of the Sensor Array and Multichannel Technical Committee of the IEEE Signal Processing Society, and the Special Area Teams ``Signal Processing for Communications and Networking'' and ``Signal Processing for Multisensor Systems'' of the EURASIP.

\end{IEEEbiography}

\end{document}